\documentclass{article}

\usepackage{PRIMEarxiv}

\usepackage[utf8]{inputenc} 
\usepackage[T1]{fontenc}    
\usepackage{hyperref}       
\usepackage{url}            
\usepackage{booktabs}       
\usepackage{amsfonts}       
\usepackage{nicefrac}       
\usepackage{microtype}      
\usepackage{lipsum}
\usepackage{amsmath}
\usepackage{fancyhdr}       
\usepackage{graphicx}       
\graphicspath{{media/}}     
\usepackage{natbib}
\usepackage{multirow}
\newcommand{\undertitle}{A Preprint}
 \newcommand\puffinlogo{\raisebox{-2pt}{\includegraphics[width=0.6em]{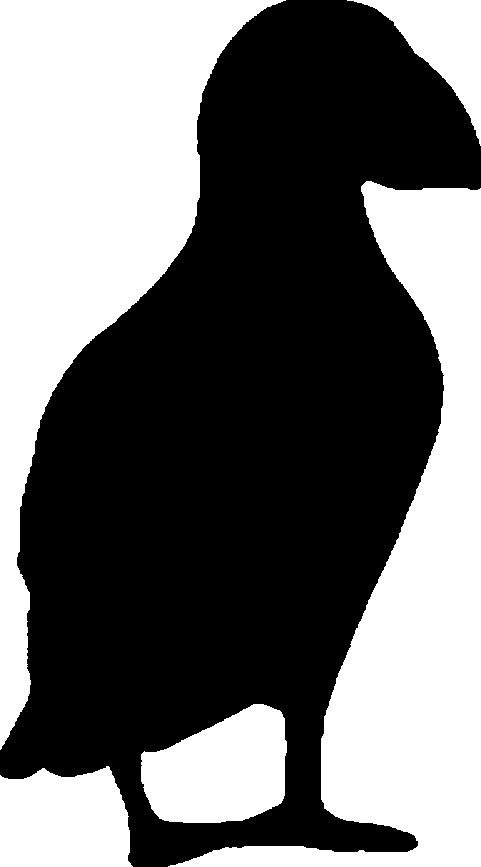}}}

\title{\puffinlogo PUFFIN: Protein Unit Discovery with Functional Supervision
}

\author{
Gökçe Uludoğan \\
Department of Computer Engineering \\
Boğaziçi University \\
Istanbul, Türkiye \\
\texttt{gokce.uludogan@bogazici.edu.tr} \\
\And
Buse Giledereli \\
Department of Computer Engineering \\
Boğaziçi University \\
Istanbul, Türkiye \\
\texttt{buse.giledereli@bogazici.edu.tr} \\
\And
Elif Ozkirimli \\
Roche Informatics \\
F. Hoffmann-La Roche AG \\
Basel, Switzerland \\
\texttt{elif.ozkirimli@roche.com} \\
\And
Arzucan Özgür \\
Department of Computer Engineering \\
Boğaziçi University \\
Istanbul, Türkiye \\
\texttt{arzucan.ozgur@bogazici.edu.tr} \\
}

\begin{document}
\maketitle

\begingroup
\renewcommand\thefootnote{}
\footnotetext{This article has been accepted for publication in Bioinformatics (ISMB 2026 Proceedings), published by Oxford University Press.}
\endgroup

\begin{abstract}
\textbf{Motivation:} 
Proteins carry out biological functions through the coordinated action of groups of residues organized into structural arrangements. These arrangements, which we refer to as protein units, exist at an intermediate scale, being larger than individual residues yet smaller than entire proteins. A deeper understanding of protein function can be achieved by identifying these units and their associations with function. However, existing approaches either focus on residue-level signals, rely on curated annotations, or segment protein structures without incorporating functional information, thereby limiting interpretable analysis of structure-function relationships. \\
\textbf{Results:} 
We introduce PUFFIN, a data-driven framework for discovering protein units by jointly learning structural partitioning and functional supervision. PUFFIN represents proteins as residue-level structure graphs and applies a graph neural network with a structure-aware pooling mechanism that partitions each protein into multi-residue units, with functional supervision that shapes the partition. 
We show that the learned units are structurally coherent, exhibit organized associations with molecular function, and show meaningful correspondence with curated InterPro annotations. Together, these results demonstrate that PUFFIN provides an interpretable framework for analyzing structure–function relationships using learned protein units and their statistical function associations.\\ 
\textbf{Availability and Implementation:} We made our source code available at \href{https:/github.com/boun-tabi-lifelu/puffin}{github.com/boun-tabi-lifelu/puffin}
.\\

\end{abstract}


\section{Introduction}
\label{sec:introduction}
Identifying the functions of proteins is essential, as these molecules serve as the primary functional units of life. While contemporary research has focused extensively on predicting entire protein function, these global models often overlook the discrete biochemical mechanisms that drive this activity. 
Structural evidence suggests that protein-level function is an emergent property arising from the coordinated behavior of groups of residues rather than isolated amino acids \citep{halabi2009protein, amitai2004network}.
Explicitly representing such subunits, therefore, provides a natural framework for linking local structural organization to global functional annotations.

We can approach this problem by drawing an analogy to natural language, namely the task of interpreting the meaning of a text written in an unfamiliar language. A text does not convey meaning as an undifferentiated whole; instead, meaning emerges from the way smaller units such as words combine within a larger context. In language, we can decipher the meaning of words by the usage of semantically similar words across texts and their association with broader themes or topics.
Similarly, protein function can be examined by relating localized regions within a structure, referred to as protein units, to protein-level functional annotation, providing a compositional perspective on how local patterns may contribute to global biological activity.

Identifying such protein units automatically would enhance functional interpretability beyond residues-level analyses, allow comparison of functional blocks across proteins, and provide a unit-level vocabulary for investigating protein function without depending on labor-intensive, expert-classified structural and functional units. Nevertheless, identifying a functionally meaningful decomposition that is consistent remains a major challenge. Existing expert-classified protein units, including domains and motifs, provide valuable biological insight but do not offer an exhaustive or unbiased decomposition of protein structure, as even comprehensive structural domain classifications cover only part of known protein structures and differ in their boundary definitions ~\citep{schaeffer_completeness_2021}, limiting their usefulness as a general framework for studying structure-function relationships at fine-grained unit-level. 

Existing approaches address parts of this problem in isolation. Decomposition methods partition proteins into structural units but 
lack functional guidance.
Sequence-based tokenization methods learn frequent motifs but ignore three-dimensional structure and functional organization \citep{suyunu_linguistic_2024}. Structure-aware discretization methods encode local geometry but typically produce small, fixed-size units optimized for reconstruction rather than function \citep{zerefa_interpretable_2025, derry_unsupervised_2025}. Complete partitioning methods assign every residue to a segment but generally yield decompositions without explicit functional guidance \citep{sangster_zero-shot_2025, sun_protein_2025}. 

In contrast, approaches focused on function prediction or interpretability incorporate supervision derived from
protein-level functional annotations but do not define structural subunits.
Recent structure-aware function prediction models integrate structural features or priors to improve protein function prediction \citep{wang_dpfunc_2025}, while interpretability analyses of protein language models identify residues or latent features associated with function \citep{nayar_paying_2025,simon_interplm_2025}. 
These approaches provide insight into functional determinants, but do not yield a structural decomposition. As a result, protein structure decomposition and function prediction are treated as separate problems, and structure–function relationships are studied at the residue level or existing expert-classified protein units such as known domains or motifs.

\begin{figure}[!t]
    \centering
    \includegraphics[width=\columnwidth]{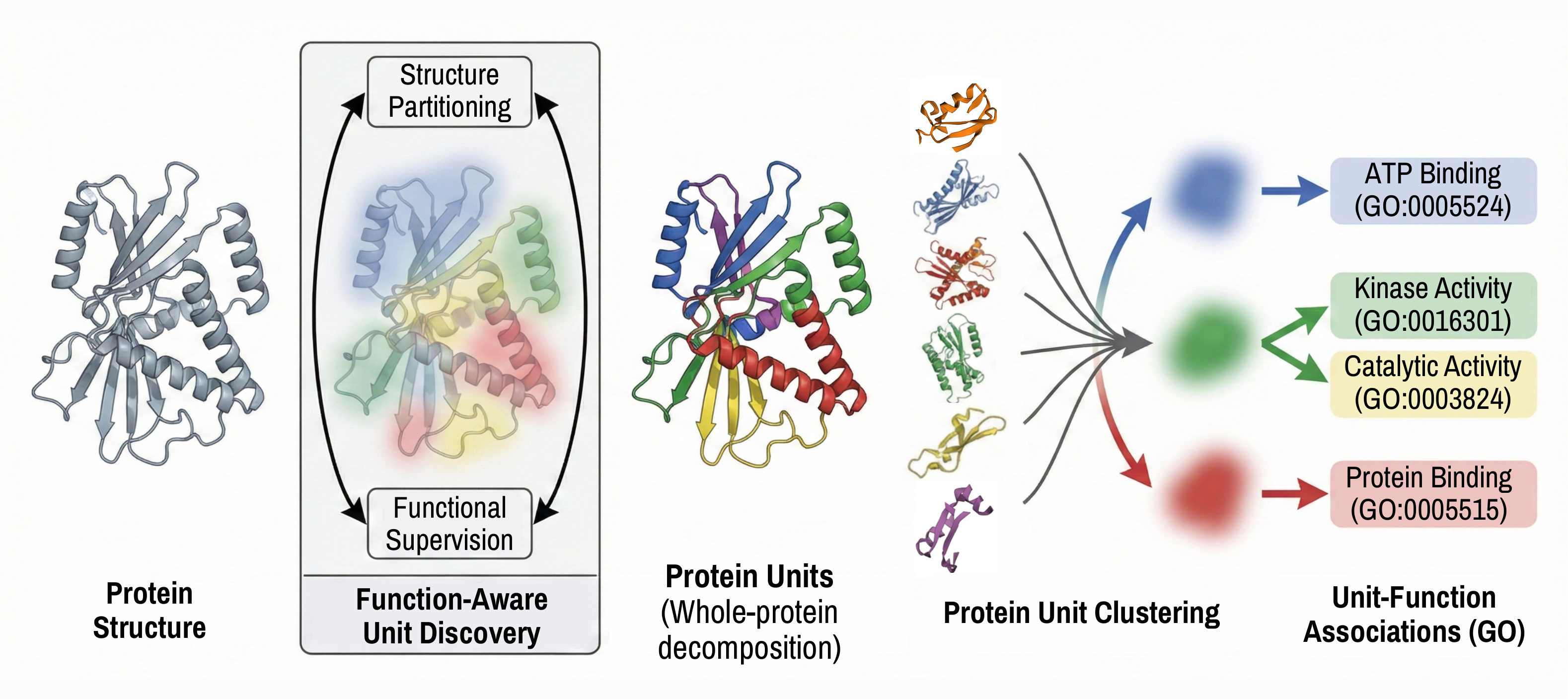}
    \caption{\textbf{Function-aware unit discovery.}
    PUFFIN jointly performs structure partitioning and protein-level functional supervision to decompose protein structures into multi-residue units. Learned units are clustered across proteins and statistically associated with Gene Ontology (GO) terms, enabling analysis of unit–function relationships.}
    \label{fig:model_goals}
\end{figure}

This motivates exploring a data-driven decomposition that
covers the entire protein, produces multi-residue units, and is
explicitly aligned with biological function. In this work,
we utilize protein-level functional annotations as
supervisory signals during training to guide the learning of
structural partitions, a setting we refer to as \textit{functional supervision}. To our knowledge, no prior method jointly learns structural partitions under such supervision. We therefore investigate whether functionally meaningful structural elements can be learned by combining structural inductive bias with protein-level functional labels, and whether the resulting units can be statistically associated with standardized functional annotations such as Gene Ontology (GO) terms \citep{ashburner_gene_2000}, as illustrated conceptually in Figure ~\ref{fig:model_goals}. 
We define units as learned, function-aware residue groupings with flexible granularity inferred directly from data under protein function supervision. 
These units group spatially proximal residues in the protein structure, without being constrained by sequence continuity. Unlike curated structural annotations (e.g., domains or binding sites), our units emerge from jointly modeling structural and functional signals. Thus, they are not constrained to predefined annotation types and can vary in size and granularity, providing a more flexible view of how functional patterns are distributed across the protein.


We introduce PUFFIN, a framework that partitions protein structures into multi-residue units using a structural-proximity-aware inductive bias and jointly learns these units with functional supervision, 
where protein-level functional annotations (in this case, GO labels) provide the learning signal that shapes the structural decomposition..
PUFFIN represents proteins as residue-level structure graphs and applies a graph neural network, particularly Graph Attention Network (GAT)~\citep{velickovic2018graph},  together with a MinCut-based pooling~\citep{bianchi2020spectral} strategy that favors units with dense internal inter-residue connectivity and weak connectivity to the remainder of the protein, to assign each residue to a unit. Unlike heuristic partitioning approaches, this objective is learned end-to-end and coupled directly to protein-level functional supervision. 
During training, unit representations refined through additional GAT layers are pooled and used to predict protein-level function, making the learned partitioning an intermediate representation through which functional supervision shapes the decomposition. The resulting units form a complete, non-overlapping decomposition of each protein into structurally coherent regions. 

To enable cross-protein analysis, we cluster unit embeddings learned from the training set and associate unit clusters with GO terms via enrichment analysis. Using a held-out set, we analyze the characteristics of the learned units based on their size, structural contiguity, and functional organization. We validate their biological relevance by comparing unit–function associations to curated InterPro annotations. Finally, we conduct case studies to further illustrate how PUFFIN identifies function-associated units in an individual protein and in a protein family. Together, these analyses establish PUFFIN as a data-driven framework for discovering function-aware protein units and analyzing unit-level structure–function relationships.
%

\section{Materials and methods}
\label{sec:methods}
\begin{figure*}[!t]
    \centering
    \includegraphics[width=\textwidth]{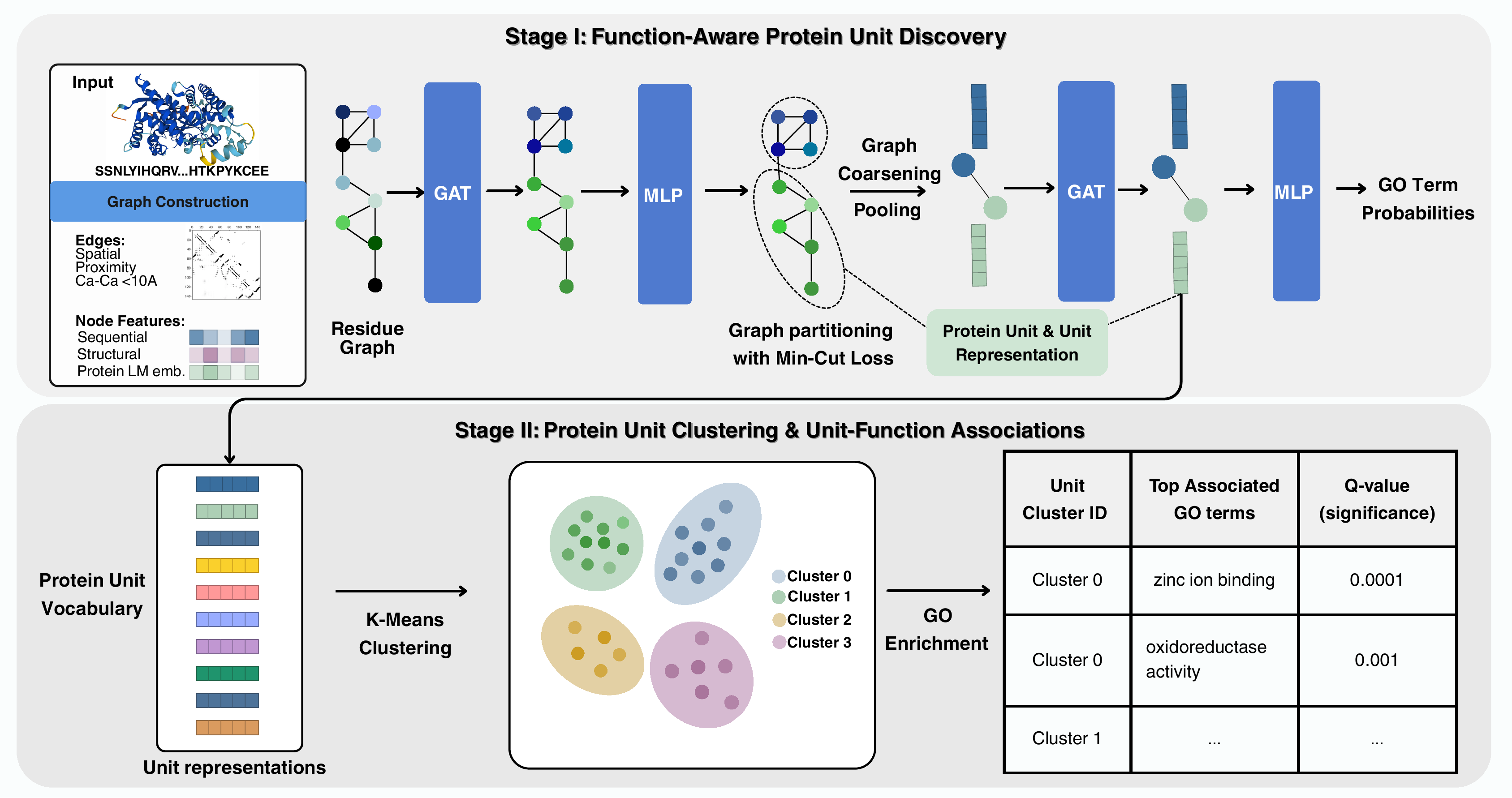}
    \caption{\textbf{Model Architecture Overview.}  
    PUFFIN processes protein structures as residue-level contact graphs with node features initialized from sequential and structural descriptors, as well as pretrained ESM-1b embeddings. During joint training (top), a residue-level Graph Attention Network (GAT) encoder produces contextualized residue representations, which are partitioned into protein units using MinCut pooling. Protein unit embeddings are refined on a coarsened graph and pooled through a unit-only readout, where a Multi-Layer Perceptron (MLP) predicts protein-level Gene Ontology (GO) terms, to ensure that functional supervision shapes partitioning. After training (bottom), protein unit embeddings from all proteins are clustered, and these clusters are analyzed via GO enrichment to associate learned units with function.
    }
    
    
    \label{fig:model_architecture}
\end{figure*}

PUFFIN is a graph-based framework for discovering function-associated protein units directly from three-dimensional protein structures. The goal is to learn a complete, structure-aware decomposition of each protein into multi-residue units whose organization is shaped jointly by structural connectivity and protein-level functional supervision. GO annotations are used only as coarse, protein-level supervision to bias unit discovery toward biologically informative regions; no residue- or region-level functional labels are used.

PUFFIN operates in three stages during training:
(i) residue-level encoding on a structure graph,
(ii) structure-aware partitioning via MinCut pooling, and
(iii) protein unit-level refinement and protein-level readout through a unit-only bottleneck.
After training, learned unit embeddings are clustered for statistical analysis of unit–function associations. 

An overview of the architecture is shown in Figure \ref{fig:model_architecture}.

\subsection{Data}

We utilise the protein structure dataset annotated with GO terms, curated by  ~\citet{gligorijevic2021structure} from the Protein Data Bank (PDB) \citep{berman2000protein}. 
Our primary focus is on the Molecular Function Ontology (MFO), as it most directly aligns with the goal of associating discovered protein units with molecular activities. 
The dataset is divided into training, validation, and test sets after filtering out obsolete structures and proteins without MF annotations, resulting in 24,936, 2,743, and 3,413 proteins, respectively.

\subsection{Residue level graph and encoding}

\subsubsection{Input Graph}
\label{subsubsec:graph}

Each protein structure is represented as a residue-level graph $G=(V,E)$, where nodes correspond to residues and edges connect residues whose $C_\alpha$ atoms are within 10~\AA. This construction yields well-connected graphs that preserve local structural interactions, consistent with prior work \citep{boadu2023combining}.
Each residue $i\in V$ is associated with an initial feature vector $\mathbf{x}_i$ comprising amino acid identity, positional encodings, backbone and side-chain torsion angles (sine/cosine). To capture rich evolutionary and biological information, we also include a representation from Evolutionary Scale Modeling (ESM),  specifically the pretrained ESM-1b embedding $\mathbf{x}_i^{\text{ESM}}$ \citep{rives2021biological}. 
We chose ESM-1b because this medium-scale model provides strong and widely used protein representations for GO prediction, while newer ESM variants have not consistently shown improved performance \citep{chen2024endowing, frolova2025mulan, vieira2025medium}.
All features are projected into a shared hidden space via a linear layer and fused to obtain the initial residue representation $\mathbf{h}_i^{(0)}$.

\subsubsection{Residue-level GAT encoder}
\label{subsubsec:gat}
Residue representations are refined using GAT layers. Given $\mathbf{h}_i^{(0)}$, each layer performs attention-weighted aggregation over spatial neighbors. We use two residue-level GAT layers with hidden dimension $H=512$.

\subsection{Unit discovery via MinCut pooling}
\label{subsubsec:mincut}

To partition residues into multi-residue units, we apply MinCut pooling to the final residue representations. A linear assignment head produces soft assignments of residues to up to $M$ units per protein. Here, 
$M$ denotes the maximum number of units the model can allocate, rather than the exact number per protein. In practice, the model partitions a protein into fewer units and adapts the segmentation to each protein. Assignment sharpness is controlled via a temperature parameter, $\tau$, which is annealed during training to encourage increasingly discrete unit assignments.

A unit corresponds to the set of residues assigned to a given unit index. We obtain discrete residue-to-unit assignments from the
soft assignment matrix $S$ by assigning each residue to the unit
with the maximum assignment probability. Units with negligible assignment mass are discarded, yielding a complete, non-overlapping partition. 
We compute soft residue-to-unit assignments using a linear head, $\mathbf{S}_{\text{logits}} = \mathbf{H}\mathbf{W}_s$ with $\mathbf{W}_s \in \mathbb{R}^{H \times M}$, 
followed by temperature-scaled softmax, 
$\mathbf{S} = \mathrm{softmax}(\mathbf{S}_{\text{logits}} / \tau) \in \mathbb{R}^{N \times M}$
where $M$ is the maximum number of units and $\tau$ controls the sharpness of assignments.

MinCut pooling is trained with the following cut objective and orthogonality regularizer:
\begin{equation}
    \mathcal{L}_c = -\frac{\operatorname{Tr}(\mathbf{S}^\top \mathbf{A} \mathbf{S})}{\operatorname{Tr}(\mathbf{S}^\top \mathbf{D} \mathbf{S})}, \quad 
    \mathcal{L}_o = \left\|\frac{\mathbf{S}^\top \mathbf{S}}{\|\mathbf{S}^\top \mathbf{S}\|_F} - \frac{\mathbf{I}_M}{\sqrt{M}}\right\|_F,
\end{equation}
where $\mathbf{A}$ is the adjacency matrix, $\mathbf{D}$ is the degree matrix, and $\mathbf{I}_M$ is the identity matrix of size $M$. The objective $\mathcal{L}_c$ favors units with dense internal and weak external connectivity in the structure graph, encouraging spatially localized residue groups, while $\mathcal{L}_o$ promotes non-overlapping, diverse unit assignments.

The final unit embeddings and coarsened graph are computed as:
\begin{equation}
    \mathbf{X}_{\text{unit}} = \mathbf{S}^\top \mathbf{H} \in \mathbb{R}^{M \times H}, \quad
    \mathbf{A}_{\text{unit}} = \mathbf{S}^\top \mathbf{A} \mathbf{S} \in \mathbb{R}^{M \times M}.
\end{equation}
Here, $\mathbf{X}_{\text{unit}}$ represents the initial unit embeddings, and $\mathbf{A}_{\text{unit}}$ is the coarsened adjacency matrix used for the subsequent GAT layers.

\subsection{Unit message passing and readout}
\label{subsubsec:segment_level}
We further refine unit representations using two GAT layers, operating on the dense coarsened graph $(\mathbf{X}_{\text{unit}},\mathbf{A}_{\text{unit}})$.
To obtain a protein embedding, we apply mean and max pooling over the set of active units and concatenate the two pooled vectors:

\begin{equation}
\mathbf{g}
=
\left[
\operatorname*{mean}_{m\in\mathcal{M}} \mathbf{x}_m
\ \big\| \
\operatorname*{max}_{m\in\mathcal{M}} \mathbf{x}_m
\right]
\in \mathbb{R}^{2H}.
\end{equation}

Finally, we project $\mathbf{g}$ to $\mathbb{R}^{H}$ using a linear layer with ReLU activation. This unit-only readout imposes an architectural bottleneck, requiring all protein-level information to pass through the learned units for functional supervision to shape segmentation. 

\subsection{Functional supervision and objective}
\label{subsubsec:func_supervision}
We use a protein-level GO prediction objective to provide functional supervision during training. 
The final protein representation $\mathbf{g}\in\mathbb{R}^{H}$ for protein $n$ is passed to a Multi-Layer Perceptron (MLP), producing logits $\hat{\mathbf{y}}_n\in\mathbb{R}^{C}$ for $C$ MF GO terms. Training uses binary cross-entropy with logits:


\begin{equation}
\mathcal{L}_{\text{func}}
=
\frac{1}{NC}
\sum_{n=1}^{N}
-\mathbf{y}_n^{\mathsf T}  \log \sigma(\hat{\mathbf{y}}_n)
-\left(\mathbf{1}-\mathbf{y}_n\right)^{\mathsf T} 
\log\!\left(1-\sigma(\hat{\mathbf{y}}_n)\right)
,
\end{equation}
where $\mathbf{y}_n\in\{0,1\}^{C}$ denotes the ground-truth label vector for protein $n$, and $\sigma(\cdot)$ is applied element-wise.

We train the model end-to-end with a joint objective:

\begin{equation}
\mathcal{L} = \mathcal{L}_{\text{func}} + \lambda_{\text{unit}}\mathcal{L}_{\text{unit}},
\end{equation}
where $\mathcal{L}_{\text{func}}$ is the multilabel classification loss for GO prediction and $\mathcal{L}_{\text{unit}}$ is the MinCut-based segmentation objective and $\lambda_{\text{unit}}$ balances structural partitioning against functional supervision.

\subsection{Protein unit clustering}
\label{subsubsec:prototypes}
\label{cluster_func_association}

After training, we extract embeddings for all active protein units within the training set. Since the joint optimization operates in a high-dimensional latent space, functionally similar units may exhibit slight embedding variances. To mitigate this variability and reduce noise in unit-level embeddings, we cluster the protein units in representation space. This clustering enables the consolidation of information across similar units and a reliable nearest-centroid mapping for unseen protein units for downstream analysis.

For each cluster $k$, we evaluate functional associations by testing each GO term $g \in \mathcal{G}_p$ for enrichment among proteins containing at least one protein unit from the cluster, relative to the remaining proteins in the training split. Enrichment significance is evaluated using a one-sided Fisher’s exact test with multiple testing correction applied via the Benjamini–Hochberg procedure. 
A cluster $k$ is considered functionally associated with all GO terms that meet a significance threshold of $q < 0.05$. For reporting and analysis, we define the set of significant associations as:

\begin{equation}
\mathcal{S}(k) = \{g \in \mathcal{G}_p \mid q_{\mathrm{Fisher}}(g) < 0.05\}
\end{equation}

\subsection{Baselines and ablation}

Our comparisons focus on \emph{protein unit discovery}. Because there is no prior method that jointly discovers protein units using both structural information and functional supervision, we adopt baselines that isolate how different sources of information shape the resulting units. In particular, we disentangle the effects of (i) functional cues from pretrained embeddings without structural constraints and (ii) structure-aware segmentation without functional supervision.

The contribution of functional cues within pretrained embeddings was evaluated using residue-level representations from ESM-1b.
For each protein, we cluster residue embeddings using $k$-means (up to 64 clusters), selecting the number of clusters by aggregating Silhouette, Calinski–Harabasz, and Davies–Bouldin scores. Cluster assignments are treated as units, and unit representations are obtained by mean-pooling the embeddings of residues assigned to each cluster. This baseline captures functional information from sequence embeddings but imposes no explicit structural constraints. To isolate the effect of functional supervision, we train a variant of PUFFIN using only the MinCut objective ($\lambda_{\text{unit}} > 0$, $\mathcal{L}_{\text{func}} = 0$). This model uses the same residue features and architecture as PUFFIN, but receives no protein-level functional training signal. Unit discovery in this setting is therefore driven solely by structural inductive biases. The units and unit representations produced by these baselines are used for comparison in subsequent analyses, and the pipeline of PUFFIN for unit-function associations is used to map these baseline units to functions in  Section~\ref{interpro2go} and Section~\ref{case_study}.

\subsection{Implementation details}

PUFFIN is implemented using PyG\footnote{\url{https://github.com/pyg-team/pytorch_geometric}}~\citep{fey2019fast} and ProteinWorkshop\footnote{\url{https://github.com/a-r-j/ProteinWorkshop}}~\citep{jamasb2023evaluating}. We employ two residue-level GAT layers with hidden dimension 512, followed by a MinCut pooling layer that produces up to $M$ units. The pooled unit representations are then processed by two additional unit-level GAT layers with the same hidden dimensionality. The function prediction head consists of a two-layer MLP with hidden dimensions $[1024, 512]$, each followed by ReLU activation and dropout with rate 0.2, and a final linear layer that outputs predictions for the 489 GO terms.

The models are trained for $20$ epochs using the Adam optimizer \citep{kingma2014adam} with a learning rate of $1e-3$, without weight decay or learning rate scheduling. We employ early stopping with a patience of 5 epochs. We use a warm-up of 3 epochs and ramp up for 2 epochs, linearly increasing the unit loss weight $\lambda_{\text{unit}}$ from 0.1 to 0.5, while annealing the MinCut assignment temperature $\tau$ from 1.0 to 0.2. The same training pipeline is applied to the MinCut-only variant, which does not include a function objective.  The balancing weight $\lambda_{\text{unit}}$ is scheduled during training, starting at $0.1$ during an initial warm-up phase and gradually increasing to $0.5$, where it is maintained for the remainder of the training.

Our objective is to learn protein units that exhibit strong associations with molecular function. We train models with maximum unit counts $M \in \{16, 32, 64\}$, chosen based on the average protein length of approximately 256 residues. 
In practice, the model uses only a subset of the available units, adapting the number of units to each protein, as illustrated by the distribution of effective unit counts for test proteins in  Figure~\ref{fig:effective}.
While function prediction performance increases with larger $M$, preliminary experiments show that increasing $M$ beyond this range results in a large number of inactive units. Therefore, we choose $M=64$, which provides sufficient capacity while avoiding excessive inactive units, as our primary goal is unit discovery rather than maximizing function prediction accuracy. 
Fixing this unit resolution, we cluster the learned units into \(K \in \{128, 256, 512, 1024, 2048, 4096\}\) clusters. Number of clusters is selected using a joint criterion that balances GO term coverage per cluster, functional enrichment within clusters, and cluster diversity, capturing a trade-off between broad functional coverage and significant, non-redundant unit cluster-function associations, with $K=1024$. All baseline methods are evaluated using the same clustering pipeline and enrichment analysis to ensure fair comparison.
Model selection is driven by unit discovery and unit–function association performance. Protein-level Molecular Function prediction performance is reported only to validate the effectiveness of the functional supervisory signal. Under this evaluation, PUFFIN achieves an \(F_{\max}\) of 0.543, an \(S_{\min}\) of 9.555, and a macro-AUPR of 0.271.

\section{Results}
\label{sec:results}
We first characterized PUFFIN units by describing their unit sizes, and internal compactness.
Next, we assessed the functional organization and specificity of individual PUFFIN units by testing whether units with similar embeddings tend to occur in proteins sharing GO annotations, and whether specific GO terms were statistically enriched among units with similar representations.
We then compared PUFFIN unit–cluster function associations against curated InterPro annotations to evaluate alignment with established functional regions across different annotation types.
Finally, we presented a qualitative case study to illustrate how PUFFIN-derived units and their associated functional annotations enable interpretable analysis of protein structure and function.

\subsection{Characterization of protein units}

\begin{figure}[!t]
    \centering
    \includegraphics[width=\linewidth]{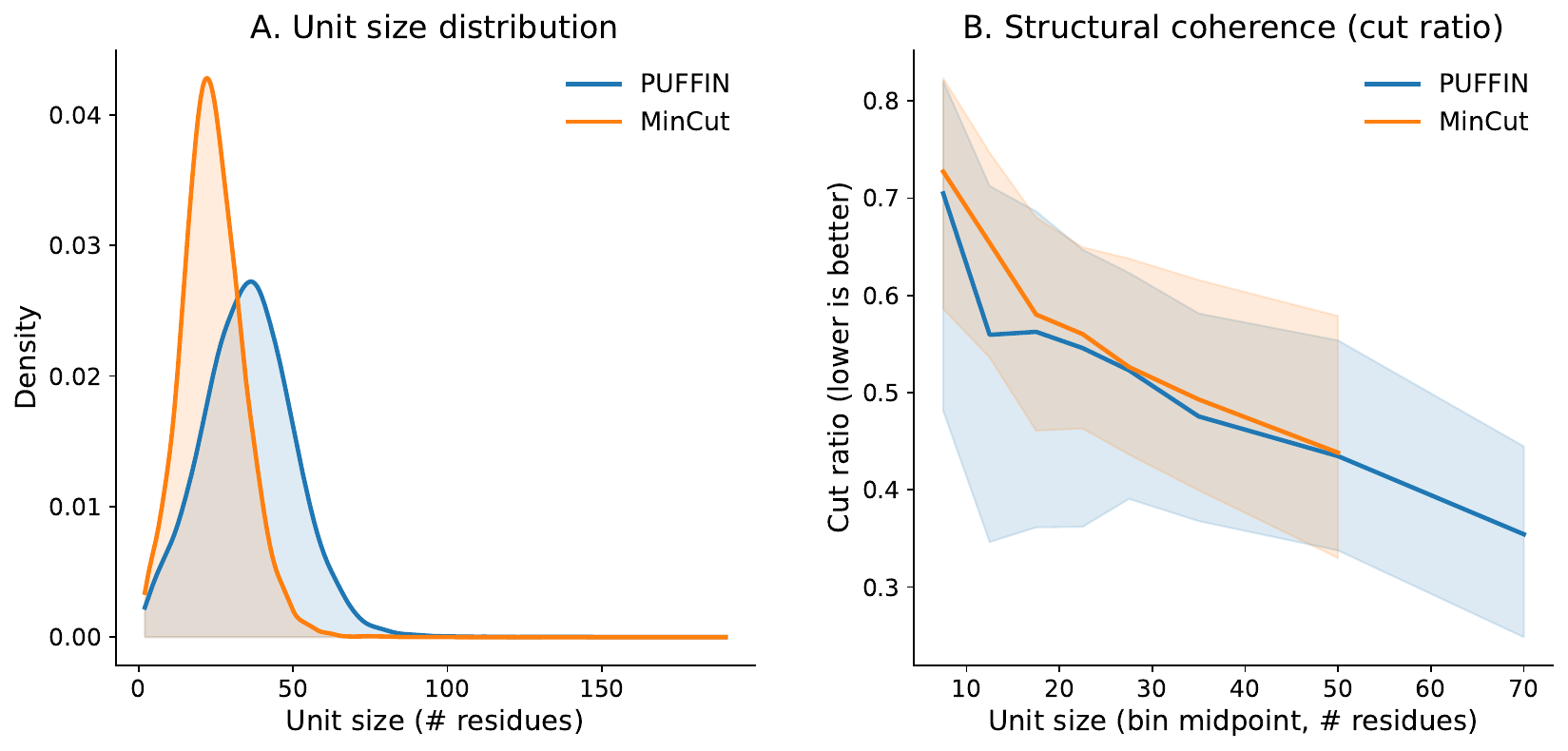}
    \caption{
    \textbf{PUFFIN unit size and structural coherence.} (A) Distribution of unit sizes shows that PUFFIN learns larger, sub-domain–scale units, relative to structure-only MinCut. (B)  The cut ratios remain comparable to MinCut across size bins, indicating that structural coherence is preserved at larger scales.
    }
    
    \label{fig:granularity}
\end{figure}

PUFFIN learned protein units with characteristic lengths in the sub-domain range.
As shown in Figure \ref{fig:granularity}A, the segment size distribution was broad, with most units spanning tens of residues rather than collapsing into very small segments.
PUFFIN units had a mean length of 35 residues, and extended up to 190 residues. Compared to structure-only MinCut, this distribution was shifted toward larger units (MinCut mean 24 residues; max 101), showing that PUFFIN favors coarser partitions.

Larger units often risk increased boundary crossings or loss of compactness. We therefore evaluated whether the larger units discovered by PUFFIN remain structurally coherent. Intra-unit compactness, measured by the mean pairwise C$\alpha$ distance within each unit, increased modestly with unit size (median 17.2~\AA{} for PUFFIN vs.\ 15.7~\AA{} for MinCut), consistent with the larger spatial extent of PUFFIN units and not indicative of structural degradation. Structural separability was assessed using the cut ratio, which measures the fraction of structural contacts crossing unit boundaries. As shown in Figure ~\ref{fig:granularity}B, PUFFIN and MinCut achieved comparable cut ratios across unit size bins. Crucially, PUFFIN more frequently produced larger units, and in this regime, cut ratios remain low, indicating that PUFFIN preserves internal connectivity rather than achieving coherence through over-fragmentation.

Overall, these results showed that PUFFIN discovered sub-domain–scale structural units with characteristic lengths of tens of residues that remain compact and well separated from the rest of the protein.

\subsection{Functional organization of units}
We next validated whether the learned representations enable meaningful association of units with molecular function.
For this analysis, we evaluated functional neighborhoods using proteins from the held-out validation split. Each protein in this split was partitioned into units, each represented by its learned embedding. All similarity searches and statistical analyses were performed using units and proteins from this split. We randomly sampled $5000$ query units from the validation set. For each query unit, we identified the $n=50$ most similar units based on cosine similarity of their learned embeddings, excluding those originating from the same protein, and examined the functional annotations of their source proteins.

\begin{figure}[!t]
    \centering
    \includegraphics[width=\linewidth]{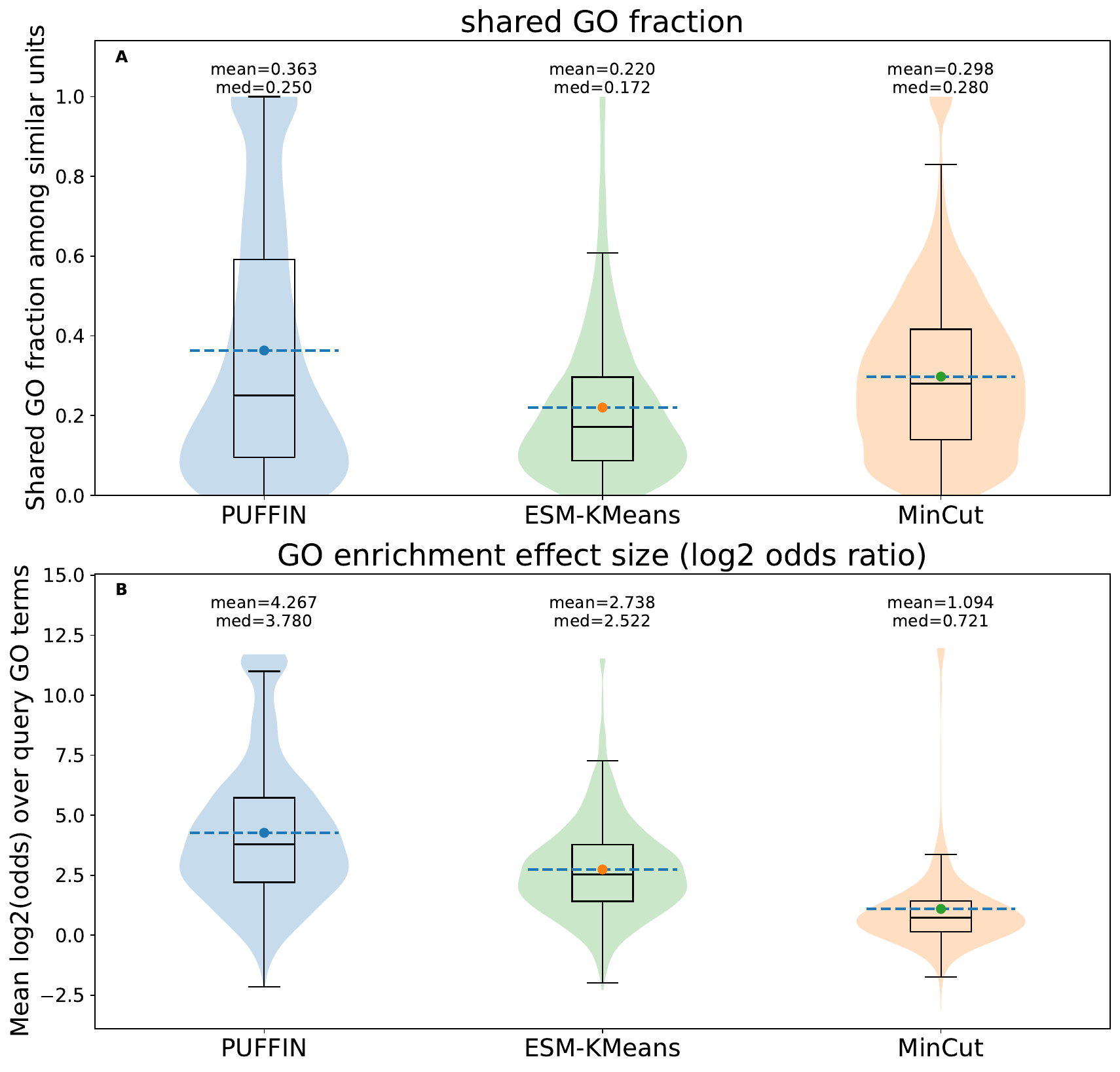}
    \caption{\textbf{
    Functional organization and specificity of learned units.} 
    (A) Units learned by PUFFIN exhibited higher shared-GO fractions than  ESM $k$-means and MinCut baselines. (B) PUFFIN units achieved stronger GO enrichment, with higher mean log$_2$ odds ratios.}
    \label{fig:neighborhood}
\end{figure}

To quantify functional consistency, we computed the shared-GO fraction, defined as the fraction of similar units whose source proteins share at least one GO term with the query protein.
This metric captures whether units with similar representations tend to occur in proteins with related functional annotations, indicating that similarity in embedding space reflects broad functional context. As shown in Figure~\ref{fig:neighborhood}A, units learned by PUFFIN exhibited higher shared-GO fractions than both baselines (mean $0.363$ vs.\ $0.220$ for ESM $k$-means and $0.298$ for MinCut). While all methods showed broad distributions, PUFFIN displayed a clear shift toward higher values, suggesting that functional supervision increases the consistency with which similar units occur in proteins sharing functional annotations.

We next assessed functional specificity using GO term enrichment within unit neighborhoods. 
For each query unit, we defined a foreground set from the source proteins of its nearest neighbors, while the remaining proteins in the split formed the background set. For each GO term annotated to the query protein, we compared its frequency in the foreground and background sets using a one-sided Fisher’s exact test. If a query protein was annotated with multiple GO terms, this resulted in one independent test per GO term for each query unit. To account for multiple testing within a given query unit, we applied Benjamini-Hochberg (BH) correction across its $|\mathcal{T}|$ term-level $p$-values, thereby controlling the false discovery rate locally for each query unit.  We summarized all term-level statistics ($p$-value, BH-corrected $q$-value, and odds ratio) at the unit level by averaging across GO terms, ensuring that each query unit contributes equally regardless of annotation count.

PUFFIN exhibits higher average BH-corrected enrichment scores (mean $-\log_{10}(q)=3.65$, median $1.46$) compared to ESM $k$-means ($3.14$, $1.52$) and MinCut ($2.64$, $0.31$), indicating stronger and more consistent statistical enrichment across queries. As shown in Figure~\ref{fig:neighborhood}B, PUFFIN units also exhibit higher enrichment effect sizes on average, with a mean log$_2$ odds ratio of $4.27$ (median $3.78$), compared to $2.74$ ($2.52$) for ESM $k$-means and $1.09$ ($0.72$) for MinCut. These results suggest that PUFFIN units are not only broadly aligned with function, but are also associated with more specific functional annotations on average.

Notably, the baselines exhibited contrasting behavior. ESM $k$-means shows lower shared-GO fractions but stronger GO enrichment than MinCut, while MinCut exhibits the opposite trend. This pattern suggests that ESM $k$-means produces units that occasionally capture function-specific signals but lack consistent organization across proteins. In contrast, MinCut yields structurally coherent units that tend to recur across proteins with broadly related functions, yet these units lack the functional specificity indicated by GO enrichment.

Together, these results showed that PUFFIN learned units such that units with similar representations are observed in proteins with related functions and selectively associated with specific molecular activities, compared to baselines, enabling unit–function analysis beyond embeddings or structural constraints alone.

\subsection{Alignment with InterPro}
\label{interpro2go}

Building on our demonstration that PUFFIN unit representations capture functional associations, we next examined how much these learned units and GO term associations align with established biological regions and their functional annotations. To address this, we utilized InterPro~\citep{mitchell_interpro_2014}, a curated database that provides residue-level functional annotation intervals ranging from localized active sites to superfamily-level classifications that may span the entire protein. For direct comparison, each InterPro annotation interval was mapped to a multilabel GO term set using InterPro2GO\footnote{
InterPro2GO mapping file, Gene Ontology Consortium.
Available at \url{https://current.geneontology.org/ontology/external2go/interpro2go}
,
version date: 2025-09-01}, thereby enabling a systematic assessment of the functional alignment between InterPro annotations, PUFFIN-derived protein units, and the baseline partitioning methodologies. 

For each protein in the test set, we identified protein units and mapped them to their nearest cluster centroids. 
We extracted all InterPro annotations for the same protein as residue-level intervals and matched them to the cluster with the highest spatial overlap, calculated as the Intersection-over-Union (IoU) between the InterPro interval and the merged residue intervals of all units in the protein assigned to that cluster. We excluded InterPro entries without GO mappings.

Functional alignment was evaluated by comparing the GO terms mapped from each InterPro annotation to the enriched GO terms of the matched cluster, ranked by their associated $q$-values (Section~\ref{cluster_func_association}), accounting for the fact that PUFFIN units often capture multiple functional roles within the same region (median: 12 GO terms per cluster). We reported standard retrieval metrics, mean reciprocal rank (MRR), hit@k, recall@k, and precision@k, computed at the level of individual InterPro annotation intervals, grouped by annotation type, aggregated per protein, and averaged across the test set. MRR and Hit@10 (H@10) are reported in Table~\ref{tab:interpro_main}, with full retrieval metrics in the Table \ref{app:tab:interpro_full}. In addition, we reported the number of proteins in the test set per category in parentheses to account for differences in the prevalence of InterPro annotation types.

As shown in Table~\ref{tab:interpro_main}, PUFFIN achieved the highest MRR and H@10 across nearly all InterPro categories, demonstrating a strong alignment with functional annotations. Improvements were consistent across both specific site annotations and broader structural classifications. Notably, while absolute performance was highest for active sites (MRR 0.6062), the most significant relative improvement occurs in conserved sites, where PUFFIN improved MRR by 67.6\% over the ESM $k$-means baseline. This suggested PUFFIN's integration of both structural and functional information allowed it to better resolve patterns that only sequence- or structure-based segmentation methods miss. 

Furthermore, retrieval performance appeared positively correlated with the functional granularity. Performance was strongest for fine-grained annotations (e.g. active site) and decreased at the family and homologous superfamily levels, where more global, protein-level functional signals are required. In the case of binding sites, ESM $k$-means narrowly led in H@10 metric, but PUFFIN maintained a significantly higher MRR (0.2927 vs 0.2242). This indicates that while both models found relevant GO terms, PUFFIN ranked them more accurately within the top results. Overall, the consistent gains across categories suggested that PUFFIN ~robustly captured functional signatures compared to baseline approaches.

\begin{table}[!htp]
\centering
\small
\caption{\textbf{InterPro retrieval performance.} Best results per type are in bold. 
Numbers in parentheses show the number of proteins that were included. 
MRR: Mean Reciprocal Rank.}
\label{tab:interpro_main}
\setlength{\tabcolsep}{6pt}
\begin{tabular}{llccc}
\toprule
InterPro type & Model & MRR & Hit@10 \\
\midrule
\multirow{3}{*}{active site (159)}
 & ESM $k$-means & 0.4518 & 0.5189 \\
 & MinCut & 0.2023 & 0.3113 \\
 & \textbf{PUFFIN} & \textbf{0.6062} & \textbf{0.6918} \\
\midrule
\multirow{3}{*}{binding site (118)}
 & ESM $k$-means & 0.2242 & \textbf{0.6737} \\
 & MinCut & 0.0896 & 0.2712 \\
 & \textbf{PUFFIN} & \textbf{0.2927} & 0.6695 \\
\midrule
\multirow{3}{*}{conserved site (408)}
 & ESM $k$-means & 0.2323 & 0.3725 \\
 & MinCut & 0.0587 & 0.1544 \\
 & \textbf{PUFFIN} & \textbf{0.3893} & \textbf{0.5233} \\
\midrule
\multirow{3}{*}{domain (1281)}
 & ESM $k$-means & 0.2705 & 0.3578 \\
 & MinCut & 0.0891 & 0.1820 \\
 & \textbf{PUFFIN} & \textbf{0.3574} & \textbf{0.4594} \\
\midrule
\multirow{3}{*}{family (1807)}
 & ESM $k$-means & 0.1833 & 0.2918 \\
 & MinCut & 0.0817 & 0.1465 \\
 & \textbf{PUFFIN} & \textbf{0.2916} & \textbf{0.4125} \\
\midrule
\multirow{3}{3cm}{homologous superfamily \newline (633)}
 & ESM $k$-means & 0.1823 & 0.2340 \\
 & MinCut & 0.0762 & 0.1313 \\
 & \textbf{PUFFIN} & \textbf{0.2597} & \textbf{0.3136} \\
\midrule
\multirow{3}{*}{overall (2741)}
 & ESM $k$-means & 0.2044 & 0.2923 \\
 & MinCut       & 0.0809 & 0.1480 \\
 & \textbf{PUFFIN} & \textbf{0.2950} & \textbf{0.4002} \\
\bottomrule
\end{tabular}

\end{table}

\begin{figure*}[!htp]
    \centering
    \includegraphics[width=\textwidth]{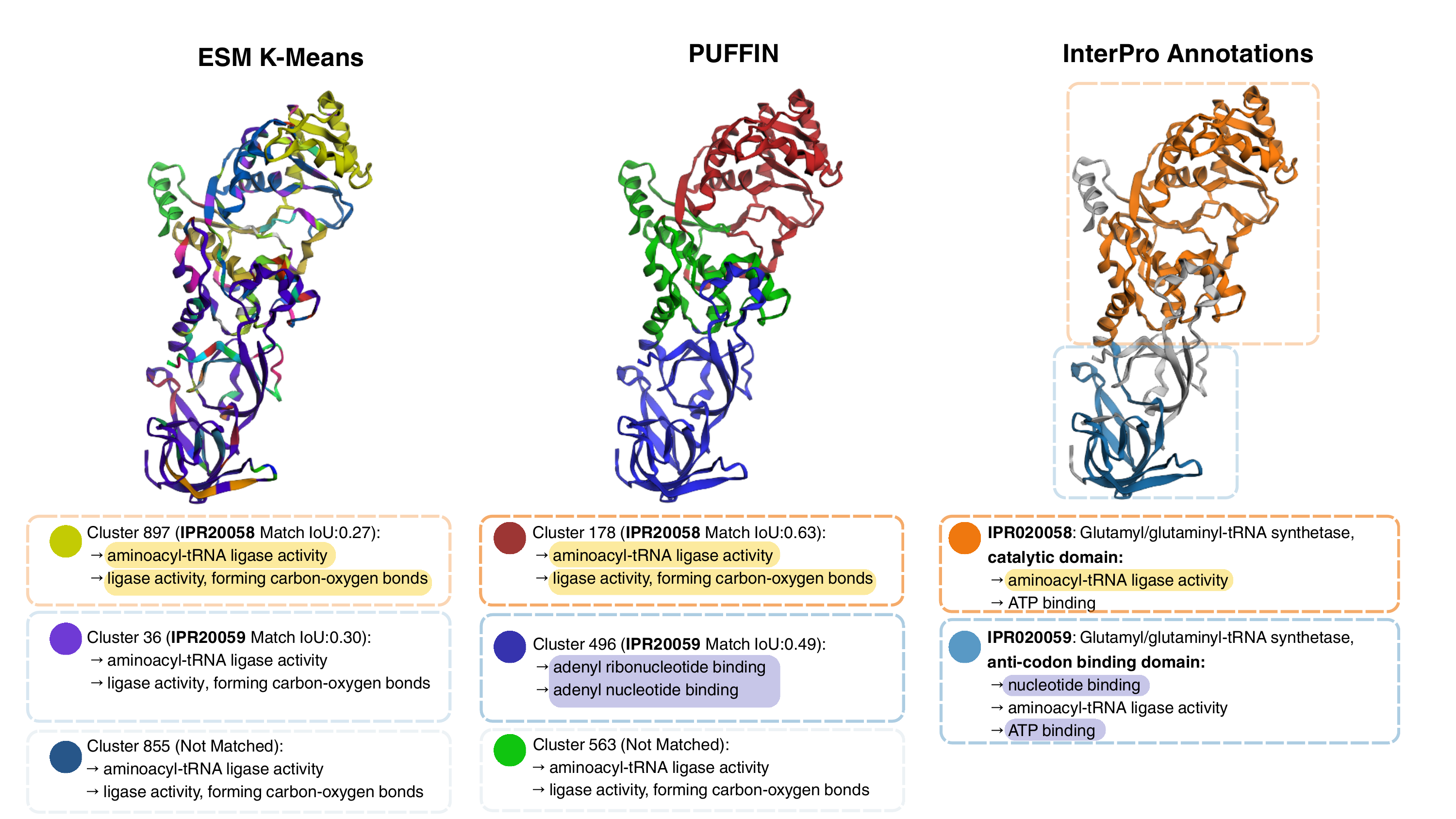}
    \caption{\textbf{Comparative partitioning and functional annotation of 2RD2, Chain A.} The protein chain is partitioned using ESM K-Means, PUFFIN, and InterPro ground-truth annotations. For each method, protein units are colored by their closest matching clusters, with their IoU score and two top-ranked enriched GO terms displayed. While both models identify the catalytic function, PUFFIN provides higher structural coherence and more specific functional labels for the anticodon-binding domain.}
    \label{fig:example_case_study}
\end{figure*}

\subsection{Case Studies}

\label{case_study}
We presented a qualitative case study illustrating how PUFFIN can be used to interpret protein structure by jointly visualizing learned protein units and their cluster-level functional annotations. We compared PUFFIN against ESM $k$-means partitioning and curated InterPro annotations.
For visualization purposes, a protein chain with non-overlapping annotations in InterPro, with existing InterPro2GO mapping was selected.
Figure \ref{fig:example_case_study} illustrates protein chain A of 2RD2, which is the Glutaminyl-tRNA synthetase mutant (C229R) from E. coli. The catalytic domain features a Rossmann fold that binds ATP, glutamine, and tRNA, while the anticodon-binding domain consists of two C-terminal $\beta$-barrels that recognize the $CUG$ anticodon \citep{2rd2_1989}. For all methods, the closest matching clusters were displayed with their associated GO terms.

Qualitatively,
ESM $k$-means tended to fragment domains into small, redundant segments, whereas PUFFIN~produced structurally coherent protein units that align more closely with InterPro domain annotations. We also observed that different clusters can exhibit similar GO term associations, which may reflect the fact that embeddings capture structural variation, such that distinct structural elements contributing to the same function form separate clusters.

For the catalytic domain, the reference InterPro GO terms are aminoacyl-tRNA ligase activity (GO:0004812) and ATP binding (GO:0005524). In alignment with this domain, the matching clusters for both methods were associated with similar GO terms, with the top two highest-ranked terms being the reference term (GO:0004812) and its parent term, ligase activity forming carbon-oxygen bonds (GO:0016875).
Although the other reference GO term, ATP  binding, did not appear among the top two terms, it was ranked sixth in both models. The higher-ranked terms mainly represented parent–child variants of the reference GO terms, indicating that both models successfully capture the functional labels associated with this domain.

While
both approaches produced comparable results for the catalytic domain, they differed noticeably in performance for the anticodon-binding domain. For this domain, InterPro assigns the GO terms nucleotide binding, aminoacyl-tRNA ligase activity, and ATP binding. While aminoacyl-tRNA ligase activity and ATP binding relate more generally to the general function of the protein, nucleotide binding is the key function, as this domain specifically mediates binding to the anticodon. In this case, the ESM $k$-means model again defaults to generic labels, such as ``aminoacyl-tRNA ligase activity." In contrast, PUFFIN ~identified the more precise terms ``adenyl ribonucleotide binding" and ``adenyl nucleotide binding." Both of these lie within two GO hierarchy steps from ``nucleotide binding", more accurately reflecting the domain’s primary role.

Overall, by learning representations informed by topological proximity and functional annotations, PUFFIN produced protein units with more coherent spatial groupings and functional assignments. In particular, by recognizing the distinct fold of the anticodon-binding domain rather than defaulting to the majority labels of the protein, PUFFIN ensured that the predicted GO terms were biologically meaningful and spatially localized.

We further illustrated the biological relevance of PUFFIN units through a case study on the Short-chain Dehydrogenase/ Reductase (SDR) protein family (Section \ref{subsec:SDR_family}). By applying hierarchical clustering to the InterPro annotations for this protein family, we identified distinct protein groups sharing similar unit cluster compositions. The conservation of several unit clusters across the family suggests that the model’s representations successfully capture recurring structural and functional elements. In particular, dominant clusters are associated with oxidoreductase activity acting on different donor groups, and the differences in unit cluster composition match these variations in donor specificity. The analysis also highlighted PUFFIN’s sensitivity to structural differences, including variations induced by ligand binding. Overall, this case study demonstrated that PUFFIN units captured functionally relevant signals and can support comparative analysis of structural and functional variation within protein families.

\section{Discussion}
\label{sec:discussion}
We introduced PUFFIN, a framework for discovering protein units by combining structure-based partitioning with protein-level functional supervision. The main idea is to decompose protein structures into multi-residue units, with their organization shaped by both structural inductive bias and functional signals. By coupling partitioning with functional supervision during training, PUFFIN provides a unified framework for studying structure-function relationships at the fine-grained unit level, 
capturing functional associations through enriched GO terms and allowing multiple functions to be associated with a unit based on enrichment strength.

Our results show that this joint formulation produces structurally coherent subdomain-scale units that remain compact and well-separated. Compared to structure-only segmentation, the learned units are larger and coarser, reflecting the influence of protein-level functional supervision.
The learned unit representations also exhibit functional organization. Compared to structure-only (MinCut) and function-only (ESM $k$-means) baselines, units learned by PUFFIN are more strongly associated with shared GO annotations and enriched GO terms, supporting unit-level analysis across proteins. Alignment with InterPro annotations further shows that these units correspond to biologically meaningful regions across annotation types. 

Case studies further illustrate the applicability of PUFFIN. A structural comparison of an individual protein showed that PUFFIN yields partitions with higher structural coherence and more specific functional annotations than baseline methods, while a family-level analysis revealed conserved unit clusters across related proteins and how their composition, together with associated GO terms, helps explain functional differences. 
More generally, PUFFIN enables the functional characterization of previously unannotated or poorly understood protein regions. By decomposing proteins with known or predicted structures into units and mapping them to learned clusters with associated GO term distributions, PUFFIN facilitates the transfer of functional signals to such regions. For example, regions overlapping Domains of Unknown Function (DUFs) can be associated with candidate functions through their corresponding unit clusters, which can serve as hypotheses for experimental validation.

While these results suggest that PUFFIN is able to identify functionally relevant protein units, the current approach has several limitations. MinCut pooling imposes a fixed maximum number of units per protein and includes an orthogonality term that encourages assignments to be evenly distributed across units, thereby limiting flexibility when proteins naturally exhibit units with varying granularity or highly uneven sizes. In addition, protein-level functional supervision provides relatively coarse guidance, potentially constraining the resolution at which finer functional distinctions can be captured. 
PUFFIN produces a non-overlapping partition of residues into distinct units, but does not explicitly model residues shared between units, such as hinge residues. While the model outputs soft assignment probabilities that could capture uncertainty near unit boundaries, residues are still assigned predominantly to a single unit. Extending the formulation to allow explicit overlapping memberships could better capture such shared roles.
Exploring more flexible unit discovery mechanisms and finer-grained functional supervision remains an open and promising direction for future work.


Overall, PUFFIN provides an interpretable, structure-guided framework for protein unit discovery and analysis. By systematically linking protein units to functional signals through unit–cluster associations, it enables data-driven exploration of protein regions, supports hypothesis generation, and offers a scalable approach for prioritizing candidate units for downstream experimental validation.

\section*{Conflict of interests}
E.O. is an employee of F. Hoffmann-La Roche Ltd.

\section*{Acknowledgments}
This work is supported by ERC grant (LifeLU, 101089287). Views and opinions expressed are however those of the author(s) only and do not necessarily reflect those of the European Union or the European Research Council Executive Agency. Neither the European Union nor the granting authority can be held responsible for them.

\bibliographystyle{abbrvnat}  
\bibliography{sections/reference}  

@inproceedings{suyunu_linguistic_2024,
	title = {Linguistic Laws Meet Protein Sequences: {A} Comparative Analysis of Subword Tokenization Methods},
	shorttitle = {Linguistic {Laws} {Meet} {Protein} {Sequences}},
	url = {},
	doi = {},
	booktitle = {2024 {IEEE} {International} {Conference} on {Bioinformatics} and {Biomedicine} ({BIBM})},
	author = {Suyunu, Burak and Taylan, Enes and Özgür, Arzucan},
	year = {2024},
}

@article{derry_unsupervised_2025,
	title = {Unsupervised learning reveals landscape of local structural motifs across protein classes},
	volume = {},
	url = {},
	doi = {},
	number = {7},
	urldate = {},
	journal = {Bioinformatics},
	author = {Derry, Alexander and Krupkin, Haim and Tartici, Alp and Altman, Russ B},
	year = {2025},}

@article{zerefa_interpretable_2025,
  title={An interpretable alphabet for local protein structure search based on amino acid neighborhoods},
  author={Zerefa, Saba and Cool, Jesse and Singh, Pramesh and Petti, Samantha},
  journal={Bioinformatics},
  volume={41},
  number={10},
  pages={btaf458},
  year={2025},
  publisher={Oxford University Press}
}

@article{sangster_zero-shot_2025,
	title = {Zero-shot segmentation using embeddings from a protein language model identifies functional regions in the human proteome},
	volume = {21},
	issn = {},
	url = {},
	doi = {},
	journal = {PLOS Computational Biology},
	author = {Sangster, Ami G. and Dufault, Cameron and Qu, Haoning and others},
	year = {2025},
	
}

@article{wang_dpfunc_2025,
	title = {{DPFunc}: accurately predicting protein function via deep learning with domain-guided structure information},
	url = {},
	doi = {},
	journal = {Nature Communications},
	author = {Wang, Wenkang and Shuai, Yunyan and Zeng, Min and others},
	year = {2025},

}

@article{simon_interplm_2025,
	title = {{InterPLM}: discovering interpretable features in protein language models via sparse autoencoders},
    author = {Simon, Elana and Zou, James},
    journal={Nature methods},
    volume={22},
    number={10},
    pages={2107--2117},
    year={2025},
    publisher={Nature Publishing Group US New York}
}

@article{nayar_paying_2025,
	title = {Paying attention to attention: {High} attention sites as indicators of protein family and function in language models},
	volume = {21},
	url = {},
	doi = {},
	journal = {PLOS Computational Biology},
	author = {Nayar, Gowri and Tartici, Alp and Altman, Russ B.},
	year = {2025},
}

@article{sun_protein_2025,
	title = {Protein {Structure} {Tokenization} via {Geometric} {Byte} {Pair} {Encoding}},
	publisher = {arXiv},
	author = {Sun, Michael and Yuan, Weize and Liu, Gang and Matusik, Wojciech and Zitnik, Marinka},
	year = {2025},
    journal={arXiv preprint arXiv:2511.11758},
}

@inproceedings{kingma2014adam,
  title={Adam: A Method for Stochastic Optimization},
  author={Kingma, Diederik and Ba, Jimmy},
  booktitle={International Conference on Learning Representations (ICLR)},
  year={2014}
}

@article{berman2000protein,
  title={The protein data bank},
  author={Berman, Helen M and Westbrook, John and Feng, Zukang and Gilliland, Gary and Bhat, Talapady N and Weissig, Helge and Shindyalov, Ilya N and Bourne, Philip E},
  journal={Nucleic acids research},
  volume={28},
  number={1},
  pages={235--242},
  year={2000},
  publisher={Oxford University Press}
}

@article{gligorijevic2021structure,
  title={Structure-based protein function prediction using graph convolutional networks},
  author={Gligorijevi{\'c}, Vladimir and Renfrew, P Douglas and Kosciolek, Tomasz and others},
  journal={Nature communications},
  volume={12},
  number={1},
  pages={3168},
  year={2021},
  publisher={Nature Publishing Group UK London}
}

@article{boadu2023combining,
  title={Combining protein sequences and structures with transformers and equivariant graph neural networks to predict protein function},
  author={Boadu, Frimpong and Cao, Hongyuan and Cheng, Jianlin},
  journal={Bioinformatics},
  volume={39},
  number={Supplement\_1},
  pages={i318--i325},
  year={2023},
  publisher={Oxford University Press}
}

@article{rives2021biological,
  title={Biological structure and function emerge from scaling unsupervised learning to 250 million protein sequences},
  author={Rives, Alexander and Meier, Joshua and Sercu, Tom  and others},
  journal={Proceedings of the National Academy of Sciences},
  volume={118},
  number={15},
  pages={e2016239118},
  year={2021},
  publisher={National Acad Sciences}
}

@article{chen2024endowing,
  title={Endowing protein language models with structural knowledge},
  author={Chen, Dexiong and Hartout, Philip and Pellizzoni, Paolo and Oliver, Carlos and Borgwardt, Karsten},
  journal={arXiv preprint arXiv:2401.14819},
  year={2024}
}

@article{frolova2025mulan,
  title={MULAN: multimodal protein language model for sequence and structure encoding},
  author={Frolova, Daria and Pak, Marina and Litvin, Anna and others},
  journal={Bioinformatics Advances},
  volume={5},
  number={1},
  pages={vbaf117},
  year={2025},
  publisher={Oxford University Press}
}

@article{vieira2025medium,
  title={Medium-sized protein language models perform well at transfer learning on realistic datasets},
  author={Vieira, Luiz C and Handojo, Morgan L and Wilke, Claus O},
  journal={Scientific Reports},
  volume={15},
  number={1},
  pages={21400},
  year={2025},
  publisher={Nature Publishing Group UK London}
}

@article{
  velickovic2018graph,
  title="{Graph Attention Networks}",
  author={Veli{\v{c}}kovi{\'{c}}, Petar and Cucurull, Guillem and Casanova, Arantxa and Romero, Adriana and Li{\`{o}}, Pietro and Bengio, Yoshua},
  journal={International Conference on Learning Representations},
  year={2018},
}

@inproceedings{bianchi2020spectral,
  title={Spectral clustering with graph neural networks for graph pooling},
  author={Bianchi, Filippo Maria and Grattarola, Daniele and Alippi, Cesare},
  booktitle={International conference on machine learning},
  year={2020},
  organization={PMLR}
}

@article{fey2019fast,
  title={Fast graph representation learning with PyTorch Geometric},
  author={Fey, Matthias and Lenssen, Jan Eric},
  journal={preprint arXiv:1903.02428},
  year={2019}
}

@inproceedings{jamasb2023evaluating,
  title={Evaluating representation learning on the protein structure universe},
  author={Jamasb, Arian R and Morehead, Alex and others},
  booktitle={NeurIPS Workshop on Machine Learning for Structural Biology},
  year={2023}
}

@article{mitchell_interpro_2014,
	title = {The {InterPro} protein families database: the classification resource after 15 years},
	volume = {43},
	number = {D1},
	journal = {Nucleic Acids Research},
	author = {Mitchell, Alex and Chang, Hsin-Yu and others},
	year = {2015},
	pages = {D213--D221},
}

@article{2rd2_1989,
author = {Mark A. Rould  and John J. Perona  and Dieter Söll  and others},
title = {Structure of \textit{E.~coli} Glutaminyl-tRNA Synthetase Complexed with tRNA$^{\mathrm{Gln}}$ and ATP at 2.8~\AA{}; Resolution},
journal = {Science},
volume = {246},
number = {4934},
pages = {1135-1142},
year = {1989}}

@article{halabi2009protein,
  title={Protein sectors: evolutionary units of three-dimensional structure},
  author={Halabi, Nizar and Rivoire, Olivier and Leibler, Stanislas and Ranganathan, Rama},
  journal={Cell},
  volume={138},
  number={4},
  pages={774--786},
  year={2009},
  publisher={Cell},
}

@article{amitai2004network,
title = {Network Analysis of Protein Structures Identifies Functional Residues},
journal = {Journal of Molecular Biology},
volume = {344},
number = {4},
pages = {1135-1146},
year = {2004},
doi = {},
author = {Gil Amitai and Arye Shemesh and others},
}

@article{schaeffer_completeness_2021,
	title = {Completeness and Consistency in Structural Domain Classifications},
	volume = {6},
	number = {24},
	journal = {ACS Omega},
	publisher = {American Chemical Society},
	author = {Schaeffer, R. Dustin and Kinch, Lisa N. and Pei, Jimin and Medvedev, Kirill E. and Grishin, Nick V.},
	year = {2021},
	pages = {15698--15707},
}

@article{ashburner_gene_2000,
	title = {Gene {Ontology}: tool for the unification of biology},
	volume = {25},
	doi = {},
	number = {1},
	journal = {Nature Genetics},
	author = {Ashburner, Michael and Ball, Catherine A. and others},
	year = {2000},
	pages = {25--29},
}

@article{kallberg_sdr,
author = {Kallberg, Yvonne and Oppermann, Udo and Jörnvall, Hans and Persson, Bengt},
title = {Short-chain dehydrogenases/reductases (SDRs)},
journal = {European Journal of Biochemistry},
volume = {269},
number = {18},
pages = {4409-4417},
doi = {https://doi.org/10.1046/j.1432-1033.2002.03130.x},
url = {https://febs.onlinelibrary.wiley.com/doi/abs/10.1046/j.1432-1033.2002.03130.x},
eprint = {https://febs.onlinelibrary.wiley.com/doi/pdf/10.1046/j.1432-1033.2002.03130.x},
year = {2002}
}

\appendix

\renewcommand{\thefigure}{S\arabic{figure}}
\renewcommand{\thetable}{S\arabic{table}}
\renewcommand{\theHfigure}{S\arabic{figure}}
\renewcommand{\theHtable}{S\arabic{table}}
\setcounter{section}{0}
\setcounter{figure}{0}
\setcounter{table}{0}

\section{Unit Cluster Construction}

After end-to-end training, we extract embeddings for all active units in the training set.
Let $\{\mathbf{e}_m\}_{m=1}^{M}$, $\mathbf{e}_m\in\mathbb{R}^{H}$, denote the resulting set of unit embeddings across proteins.

Unit embeddings are $\ell_2$-normalized and transformed using a debiasing procedure fitted on the training set:
\begin{equation}
\mathbf{e}'_m=\operatorname{norm}\!\left(\mathbf{e}_m-\boldsymbol{\mu}-\sum_{r=1}^{R}
\left\langle \mathbf{e}_m - \boldsymbol{\mu}, \mathbf{p}_r \right\rangle\mathbf{p}_r\right),
\end{equation}
where $\boldsymbol{\mu}$ is the mean embedding and $\{\mathbf{p}_r\}_{r=1}^{R}$ are the top principal components of centered training embeddings.
The same transform is applied to validation and test units.

We learn $P$ global clusters by clustering the transformed training embeddings $\{\mathbf{e}'_m\}$ using spherical $k$-means, yielding unit-norm centroids $\{\mathbf{c}_p\}_{p=1}^{P}$.
Each unit embedding is assigned to its nearest prototype using cosine similarity.

Cluster learning is performed using spherical $k$-means on the unit embeddings and used $n_{\text{init}}=5$ random initializations and 40 update iterations. We explore different number of clusters  $P \in \{128, 256, 512, 1024, 2048, 4096\}$. The optimal value of $P$ is selected using a joint scoring criterion that balances prototype coverage and functional specificity, detailed in Section \ref{subsec:prototype_choose}.

\section{Evaluation}

The evaluation of PUFFIN involves several aspects: i) characterization of units, ii) functional analysis of units and their corresponding embeddings, iii) assessment of its function prediction performance. Additionally, unit clusters are characterised by their usage and functional enrichment.

\subsection{Protein unit evaluation}
\label{sec:segment_eval}
We assess protein units identified by PUFFIN along two complementary axes: (i) their granularity, sequence contiguity, and structural compactness and separation and (ii) the functional coherence of their learned embeddings with respect to GO annotations.

\paragraph{Characterization metrics.}


Let a protein consist of $N$ residues indexed by $i=1,\dots,N$, with hard unit assignments
\[
a_i = \arg\max_{m\in\{1,\dots,M\}} s_{im},
\]
and let $M$ denote the set of active unit defined previously.
For a given protein, the residue set of unit $m$ is
\[
\mathcal{R}_m = \{\, i \in \{1,\dots,N\} \mid a_i = m \,\}, 
\qquad m \in M.
\]

The size of unit $m$ is defined as
\[
n_{\mathrm{res}}(m) = |\mathcal{R}_m|.
\]
\textbf{Sequential coherence.}
To quantify fragmentation in sequence order, we define the number of contiguous
runs of unit $m$ as
\[
r_m
=
1
+
\sum_{i=2}^{N}
\mathbb{I}\!\left[a_i = m \;\wedge\; a_{i-1} \neq m \right],
\]
where $\mathbb{I}[\cdot]$ denotes the indicator function.
We report the normalized fragmentation ratio
\[
\mathrm{frag}(m) = \frac{r_m}{n_{\mathrm{res}}(m)},
\]
with lower values indicating higher sequence contiguity.

\textbf{Structural compactness and connectivity.}
Let $\mathbf{x}_i \in \mathbb{R}^3$ denote the C$\alpha$ coordinate of residue $i$,
and let $\mathcal{C}$ denote the set of residue--residue contacts.

For each unit $m$, the mean intra-unit C$\alpha$ distance is computed as
\[
d_{\mathrm{intra}}(m)
=
\frac{1}{|\mathcal{R}_m|(|\mathcal{R}_m|-1)}
\sum_{\substack{i,j \in \mathcal{R}_m \\ i \neq j}}
\|\mathbf{x}_i - \mathbf{x}_j\|_2.
\]

We further define the cut ratio of unit $m$ as the fraction of contacts crossing
the unit boundary:
\[
\mathrm{cut}(m)
=
\frac{
\bigl|\{(i,j)\in\mathcal{C} \mid a_i = m,\; a_j \neq m\}\bigr|
}{
\bigl|\{(i,j)\in\mathcal{C} \mid a_i = m\}\bigr|
}.
\]

\paragraph{Functional coherence evaluation}
We evaluate whether learned unit embeddings capture functionally meaningful patterns using a $k$-nearest neighbor (kNN) analysis in embedding space.
Each active unit $m$ from protein $p$ is associated with an embedding $\mathbf{z}_{p,m} \in \mathbb{R}^d$.
For each unit embedding, we retrieve its neighborhood
$\mathcal{N}_m$ consisting of the $k$ nearest unit embeddings under cosine similarity.
Each neighbor $j \in \mathcal{N}_m$ originates from a protein denoted $p(j)$.

Let $\mathcal{G}_p$ denote the set of Gene Ontology (GO) terms annotating protein $p$.
We quantify functional coherence of unit neighborhoods using the following metrics.

\textbf{Neighborhood overlap.}
Neighborhood overlap measures the fraction of unit proteins sharing at least one GO term with the query protein:
\[
\mathrm{overlap}(m)
=
\frac{1}{|\mathcal{N}_m|}
\sum_{j \in \mathcal{N}_m}
\mathbb{I}\!\left(
\mathcal{G}_{p(j)} \cap \mathcal{G}_p \neq \emptyset
\right).
\]
Higher values indicate neighborhoods enriched for functionally related proteins.

\textbf{GO term enrichment.}
To assess finer-grained functional specificity, we test each GO term $g \in \mathcal{G}_p$ for over-representation among neighbor proteins relative to the background protein set of the evaluation split.
For each term, enrichment significance is assessed using a one-sided Fisher exact test.
For each unit $m$, we report the most significant enrichment scores among the query protein’s terms:
\[
p^{\ast}(m) = \min_{g \in \mathcal{G}_p} p_{\mathrm{Fisher}}(g).
\]


\paragraph{Cluster evaluation} 
We further evaluate learned unit clusters.
Each unit is assigned to its nearest cluster embedding, and cluster usage is characterized by the distribution of assigned units across proteins.

We perform functional enrichment analysis at the cluster level by testing GO term over-representation among proteins contributing units to each prototype, using a Fisher exact test analogous to the unit neighborhood analysis.

Finally, we assess the correspondence between discovered units and known functional annotations by comparing prototype-associated units to InterPro regions.
InterPro annotations are mapped to GO terms via InterPro2GO. We rank clusters for each annotation based on GO term enrichment, and report standard retrieval metrics including mean reciprocal rank (MRR), Hit Rate@K, precision, and recall.

\subsection{Function prediction metrics}

We evaluate function prediction performance using the standard metrics: the maximum F-measure ($F_{max}$), and the area under the precision-recall curve (AUPRC). 

The maximum F-measure ($F_{max}$) is calculated as the maximum harmonic mean of precision and recall across different confidence thresholds $t$:
\begin{equation}
F_{\max} = \max_{t} \left\{ \frac{2 \cdot pr(t) \cdot rc(t)}{pr(t) + rc(t)} \right\}
\end{equation}
where precision $pr(t)$ and recall $rc(t)$ at threshold $t$ are defined as:
\begin{align}
pr(t) &= \frac{1}{m(t)} \sum_{i=1}^{m(t)} \frac{|P_i(t) \cap T_i|}{|P_i(t)|} \\
rc(t) &= \frac{1}{n} \sum_{i=1}^{n} \frac{|P_i(t) \cap T_i|}{|T_i|}
\end{align}
where $P_i(t)$ represents the set of predicted GO terms with confidence score $\geq t$ for protein $i$, $T_i$ is the set of true GO terms for protein $i$, $n$ is the number of proteins with at least one GO term annotation, and $m(t)$ is the number of proteins with at least one prediction at threshold $t$.

For function-centric evaluation, we employ a macro-averaging approach across all GO terms, calculating  the Area Under the Precision-Recall Curve (AUPRC)  for each term separately:
\begin{equation}
AUPRC_j = \sum_{k=1}^{n-1} (r_{j,k+1} - r_{j,k}) \cdot \frac{p_{j,k+1} + p_{j,k}}{2}
\end{equation}
where $p_{j,k}$ and $r_{j,k}$ are the precision and recall values at the $k$-th threshold for GO term $j$. The macro-averaged metrics are then computed as:
\begin{equation}
M-AUPRC = \frac{1}{|J|}\sum_{j=1}^{|J|} AUPRC_j
\end{equation}
where $|J|$ is the total number of GO terms. 


\subsection{Cluster evaluation}

We evaluate cluster quality by examining the enriched GO terms associated with each cluster after clustering and then computing coverage, functional enrichment, and cluster diversity for them.

Let $K$ denote the number of clusters, $\mathcal{C} = \{C_1, \dots, C_K\}$ the set of all units mapping to each cluster, $\mathcal{G}_{all}$ the set of all GO terms observed in the training set, and the function $\mathcal{S}(k)$ returning the set of all enriched GO terms for cluster k.




\subsubsection{GO Coverage}
GO coverage measures the fraction of enriched GO terms that are represented by at least one cluster:
\begin{equation}
GC(K) =
\frac{\left| \bigcup_{i=1}^{K} \mathcal{S}(C_i) \right|}{|\mathcal{G}_{all}|},
\end{equation}
where $\mathcal{G}(C_i)$ denotes the set of enriched GO terms associated with the cluster $C_i$.
Higher values indicate broader functional coverage.

\subsubsection{GO per cluster}
This metric computes the mean number of enriched GO terms per cluster:
\begin{equation}
GpC(K) =
\frac{1}{K} \sum_{i=1}^{K} |\mathcal{S}(C_i)|.
\end{equation}
Lower values indicate more functionally specific clusters.

\subsubsection{Clusters per GO}
Clusters per GO measures the average number of clusters for which a GO term is enriched:
\begin{equation}
CpG(K) =
\frac{1}{|\mathcal{G}_{all}|}
\sum_{g \in \mathcal{G}_{all}}
\left| \{ C_i \mid g \in \mathcal{S}(C_i) \} \right|.
\end{equation}
Lower values reflect reduced functional fragmentation across clusters.

\subsubsection{Enriched Cluster Fraction (\texorpdfstring{$N/K$}{N/K})}
The fraction of enriched clusters is then defined as: 

\begin{equation}
ECF(K) =
\frac{\left| \{ C_i \mid |\mathcal{S}(C_i)| > 0 \} \right|}{K}
\end{equation}

where $N$ is the number of clusters containing at least one enriched GO term. Values close to 1 indicate that most clusters are functionally meaningful.

\subsubsection{Joint Score}
The joint score combines coverage, specificity, and enrichment consistency into a single scalar objective:

\begin{equation}
\label{eq:joint}
\mathrm{Joint}(K) =
\frac{
GC(K)\,\log\!\left(1 + ECF(K)\right)
}{
\log\!\left(1 + GpC(K)\right)^2\,
\log\!\left(1 + CpG(K)\right)
},
\end{equation}
where $GpC(K)$ is more hardly penalised as functionally specific clusters are more important than redundancy.

\section{Hyperparameter selection}

\subsection{Unit Size}
We explore models with varying maximum numbers of units per protein, denoted by $M \in \{16, 32, 64\}$. We evaluate how unit properties change as $M$ increases, focusing on diversity, fragmentation, structural coherence, compactness, and size-related characteristics.

{Figure~\ref{fig:effective} clarifies the role of $M$, showing the effective number of units used when segmenting test proteins. Although $M$ defines the maximum number of units that can be allocated, the model  partitions a protein into fewer units and adapts the segmentation to each protein in practice.

Figure~\ref{fig:segment} summarizes the results for MinCut-based segmentation and PUFFIN. Both methods show similar qualitative trends across  $M$. However, PUFFIN consistently favors functionally coherent segmentation, often at the expense of stricter structural constraints.

At lower values of $M$, both methods achieve high segment diversity, while a decrease is observed at $M=64$, more pronounced for PUFFIN. This suggests increasing specialization of units as the number of units per protein grows.

Fragmentation behavior is comparable between methods at $M=16$ and 
$M=32$. At $M=64$, PUFFIN exhibits lower fragmentation than MinCut, indicating more contiguous units along the sequence at higher granularity.

Structural coherence improves for both methods as 
$M$ increases, with PUFFIN showing a larger improvement at $M=64$. In contrast, structural compactness exhibits method-specific behavior, where PUFFIN forms larger and less compact segments at higher $K$, while MinCut maintains more compact units.

Consistent with these trends, PUFFIN produces larger units with longer sequence spans at $M=64$ compared to MinCut, reflecting a preference for fewer, longer units rather than uniformly small fragments.

\begin{figure*}[!ht]
    \centering
    \includegraphics[width=0.5\linewidth]{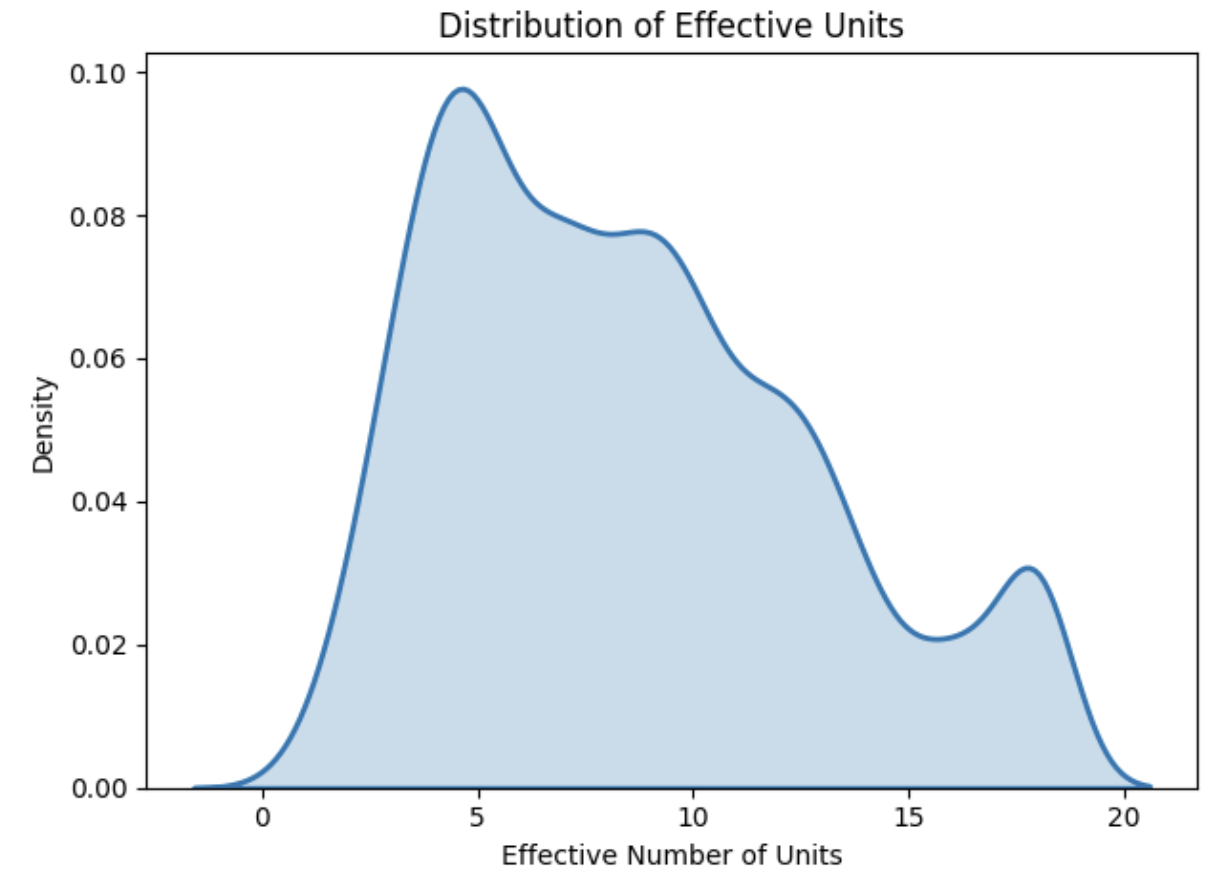}
    \caption{Effective number of units used in segmenting test proteins.}
    \label{fig:effective}
\end{figure*}

\begin{figure*}[!ht]
    \centering
    \includegraphics[width=\linewidth]{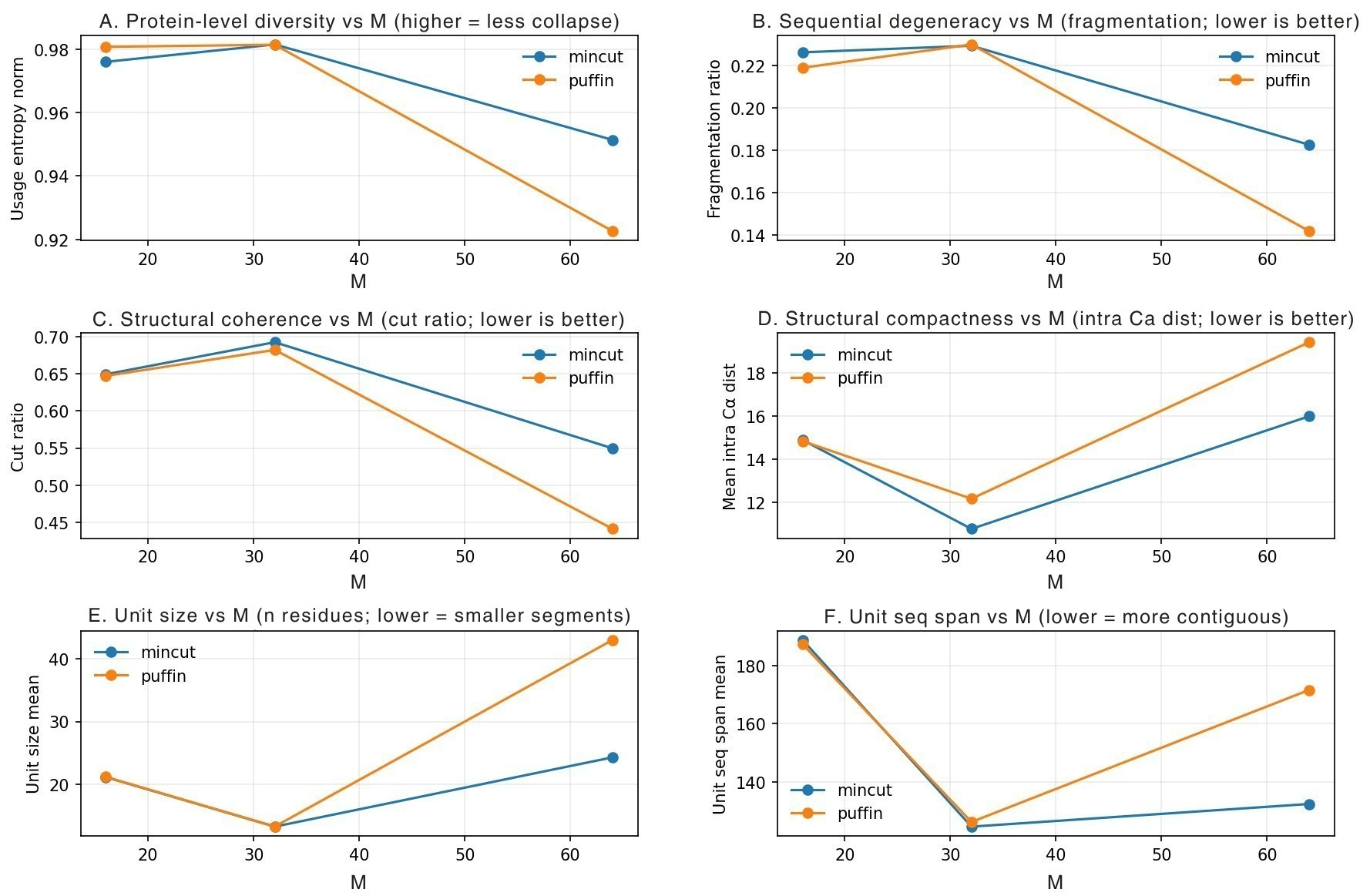}
    \caption{Unit characterization.}
    \label{fig:segment}
\end{figure*}

\subsection{Cluster Size}
\label{subsec:prototype_choose}

We select the cluster size by jointly considering coverage, functional enrichment, and cluster diversity using Equation~\ref{eq:joint}. Figure~\ref{fig:prototype_k_selection} shows how these criteria change as the number of prototypes increases.

We identify the $K$ value, 1024, that balances coverage, specificity, and robustness of functional annotation.

GO coverage stays roughly constant as the number of clusters increases, while enrichment-related measures, especially the fraction of enriched clusters, degrade at large cluster sizes. At the same time, larger cluster sets reduce diversity within clusters but increase cluster redundancy.

The joint score peaks at an intermediate cluster size, reflecting a trade-off between broad coverage and meaningful functional enrichment. Based on this analysis, we select the cluster size corresponding to the maximum joint score.

\begin{figure*}[!t]
    \centering
    \includegraphics[width=\textwidth]{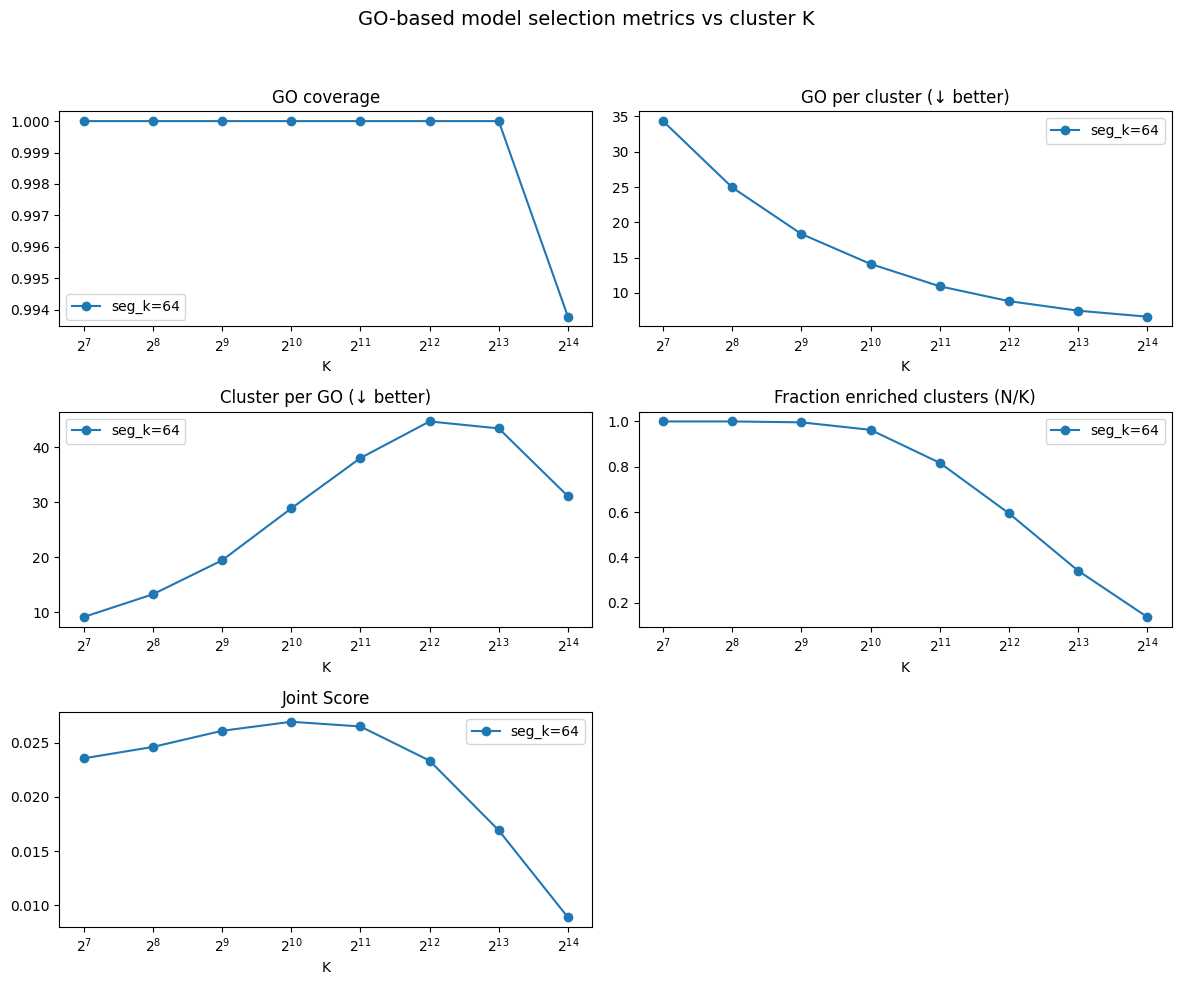}
    \caption{Cluster k was selected by observing GO enrichment metrics.}
    \label{fig:prototype_k_selection}
\end{figure*}

\subsection{InterPro retrieval performance}

Comprehensive retrieval metrics are reported to enable comparison with InterPro annotations across the various annotation categories (Table~\ref{app:tab:interpro_full}). Since a single InterPro annotation may map to multiple GO terms, the evaluation is performed in a multilabel setting, with the InterPro2GO mapping as the reference label set. 


Hit@k measures whether at least one reference GO term appears among the top-k enriched GO terms for the matched prototype. Precision@k is defined as the fraction of the top-k predicted GO terms that are present in the reference set, while Recall@k measures the fraction of reference GO terms that are recovered within the top-k predictions. In addition, mean reciprocal rank (MRR) is reported to capture the rank of the first correctly recovered GO term.

\begin{table*}[!p]
\caption{Comprehensive InterPro retrieval performance metrics across all categories. Best results per category and metric are in bold. Values in parentheses next to the InterPro type indicate the number of proteins evaluated. MRR: Mean Reciprocal Rank, H@k: Hit at k, P@k: Precision at k and R@k: Recall at k.}
\label{app:tab:interpro_full}
\centering
\small
\setlength{\tabcolsep}{5pt} 
\begin{tabular}{llccccccccc}
\toprule
\textbf{InterPro Type} & \textbf{Model} & \textbf{H@1} & \textbf{H@3} & \textbf{H@5} & \textbf{P@1} & \textbf{P@3} & \textbf{P@5} & \textbf{R@3} & \textbf{R@5} & \textbf{R@10} \\
\midrule
\multirow{3}{*}{active\_site (159)} & ESM-$k$means & 0.3931 & 0.5063 & 0.5063 & 0.3931 & 0.1688 & 0.1013 & 0.5063 & 0.5063 & 0.5189 \\
 & MinCut & 0.1509 & 0.2390 & 0.2579 & 0.1509 & 0.0797 & 0.0516 & 0.2358 & 0.2547 & 0.3113 \\
 & \textbf{PUFFIN} & \textbf{0.5723} & \textbf{0.6226} & \textbf{0.6384} & \textbf{0.5723} & \textbf{0.2138} & \textbf{0.1314} & \textbf{0.6226} & \textbf{0.6384} & \textbf{0.6887} \\
\midrule
\multirow{3}{*}{binding\_site (118)} & ESM-$k$means & 0.0847 & 0.1441 & \textbf{0.6271} & 0.0847 & 0.0508 & \textbf{0.1271} & 0.1059 & \textbf{0.5890} & \textbf{0.6419} \\
 & MinCut & 0.0339 & 0.0466 & 0.2246 & 0.0339 & 0.0155 & 0.0449 & 0.0360 & 0.2097 & 0.2535 \\
 & \textbf{PUFFIN} & \textbf{0.1780} & \textbf{0.2373} & 0.6017 & \textbf{0.1780} & \textbf{0.0819} & 0.1220 & \textbf{0.1822} & 0.5424 & 0.6017 \\
\midrule
\multirow{3}{*}{conserved\_site (408)} & ESM-$k$means & 0.1785 & 0.2574 & 0.2949 & 0.1785 & 0.1070 & 0.0776 & 0.1935 & 0.2326 & 0.3015 \\
 & MinCut & 0.0188 & 0.0654 & 0.0923 & 0.0188 & 0.0218 & 0.0185 & 0.0487 & 0.0675 & 0.1023 \\
 & \textbf{PUFFIN} & \textbf{0.3297} & \textbf{0.4105} & \textbf{0.4939} & \textbf{0.3297} & \textbf{0.1679} & \textbf{0.1262} & \textbf{0.3321} & \textbf{0.4007} & \textbf{0.4412} \\
\midrule
\multirow{3}{*}{domain (1281)} & ESM-$k$means & 0.2284 & 0.2856 & 0.3178 & 0.2284 & 0.0970 & 0.0682 & 0.2351 & 0.2701 & 0.3280 \\
 & MinCut & 0.0476 & 0.1002 & 0.1272 & 0.0476 & 0.0341 & 0.0268 & 0.0836 & 0.1080 & 0.1591 \\
 & \textbf{PUFFIN} & \textbf{0.3094} & \textbf{0.3836} & \textbf{0.4224} & \textbf{0.3094} & \textbf{0.1326} & \textbf{0.0951} & \textbf{0.3354} & \textbf{0.3834} & \textbf{0.4201} \\
\midrule
\multirow{3}{*}{family (1807)} & ESM-$k$means & 0.1361 & 0.2055 & 0.2422 & 0.1361 & 0.0783 & 0.0586 & 0.1719 & 0.2067 & 0.2468 \\
 & MinCut & 0.0520 & 0.0924 & 0.1084 & 0.0520 & 0.0329 & 0.0245 & 0.0776 & 0.0921 & 0.1234 \\
 & \textbf{PUFFIN} & \textbf{0.2390} & \textbf{0.3080} & \textbf{0.3635} & \textbf{0.2390} & \textbf{0.1152} & \textbf{0.0855} & \textbf{0.2538} & \textbf{0.3070} & \textbf{0.3551} \\
\midrule
\multirow{3}{*}{homologous\_sf (633)} & ESM-$k$means & 0.1590 & 0.1958 & 0.2101 & 0.1590 & 0.0887 & 0.0639 & 0.1645 & 0.1905 & 0.2135 \\
 & MinCut & 0.0500 & 0.0895 & 0.1014 & 0.0500 & 0.0363 & 0.0259 & 0.0775 & 0.0917 & 0.1185 \\
 & \textbf{PUFFIN} & \textbf{0.2310} & \textbf{0.2679} & \textbf{0.2995} & \textbf{0.2310} & \textbf{0.1268} & \textbf{0.0919} & \textbf{0.2433} & \textbf{0.2831} & \textbf{0.2953} \\
\midrule
\multirow{3}{*}{repeat (111)} & ESM-$k$means & \textbf{0.0090} & \textbf{0.1059} & \textbf{0.1059} & \textbf{0.0090} & \textbf{0.0353} & \textbf{0.0212} & \textbf{0.0514} & \textbf{0.0514} & \textbf{0.0634} \\
 & MinCut & 0.0000 & 0.0045 & 0.0045 & 0.0000 & 0.0015 & 0.0009 & 0.0023 & 0.0023 & 0.0023 \\
 & \textbf{PUFFIN} & 0.0000 & 0.0090 & 0.0315 & 0.0000 & 0.0030 & 0.0063 & 0.0030 & 0.0143 & 0.0315 \\
\midrule
\multirow{3}{*}{overall (2494)} & ESM-$k$means & 0.1649 & 0.2204 & 0.2548 & 0.1649 & 0.0810 & 0.0594 & 0.1875 & 0.2237 & 0.2611 \\
 & MinCut & 0.0519 & 0.0873 & 0.1065 & 0.0519 & 0.0306 & 0.0232 & 0.0756 & 0.0935 & 0.1292 \\
 & \textbf{PUFFIN} & \textbf{0.2475} & \textbf{0.3137} & \textbf{0.3634} & \textbf{0.2475} & \textbf{0.1154} & \textbf{0.0846} & \textbf{0.2714} & \textbf{0.3215} & \textbf{0.4002} \\
\bottomrule
\end{tabular}

\end{table*}

\subsection{Case Study on Short-chain Dehydrogenase/Reductase (SDR) Protein Family }
\label{subsec:SDR_family}

\begin{figure*}[t]
    \centering
    \includegraphics[width=0.9\textwidth]{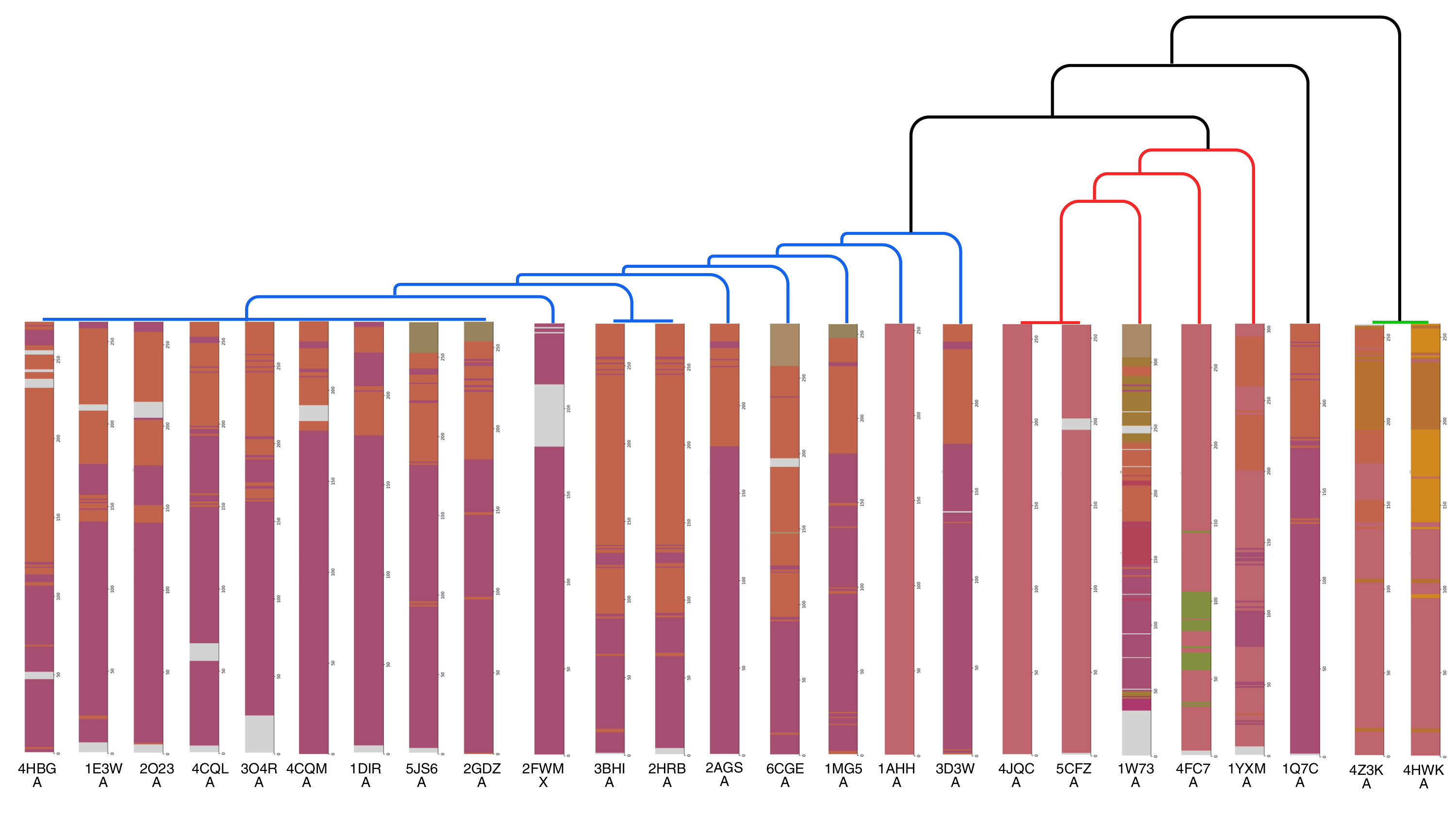}
    \caption{\textbf{Hierarchical Clustering and Sequence Segmentation Maps of the SDR Protein Family.} The dendrogram (top) illustrates the relationships between various SDR proteins based on their \textbf{InterPro accession-based clusters derived with Jaccard scores}. The branching colors (blue, red, and green) indicate distinct sub-groups identified. Below each leaf of the dendrogram is a \textbf{protein sequence unit cluster visualisation} for a specific protein. These vertical bars represent the residue index, with different colors representing their corresponding PUFFIN-based unit clusters.  }
    \label{fig:family_case_study}
\end{figure*}

To assess whether PUFFIN-derived unit clusters are conserved among protein families, we performed an analysis on Short-chain Dehydrogenase/Reductase (SDR) protein family (IPR002347) Figure \ref{fig:family_case_study}.  The SDR family is a large family of enzymes most of which function as oxidoreductases. These proteins display high sequence diversity and perform a vast range of metabolic roles, making them an ideal family for an analysis of the PUFFIN-derived units.  Our analysis included all 25 SDR-labeled proteins in the test set, without any exclusions. 

\subsubsection{Unit Cluster Visualisation on Protein Sequence}

{To visualize high-dimensional unit representations, we implemented a deterministic mapping of unit cluster centroids into the RGB color space. Centroid embeddings were projected onto their first three principal components using Principal Component Analysis (PCA) to capture primary structural variance. These components were then globally min-max normalized to a $[0, 1]$ range.
\begin{equation}
C_{norm} = \frac{C_{pca} - \min(C_{pca})}{\max(C_{pca}) - \min(C_{pca}) + \epsilon}
\end{equation}

The resulting values were mapped to red, green, and blue channels respectively, and converted to hexadecimal codes. This ensures that unit clusters with similar features are represented by perceptually similar colors across all proteins. 

\subsubsection{Comparison with InterPro Annotations}

To compare the unit clusters assigned by PUFFIN with annotations from InterPro, we conducted a  \textit{hierarchical clustering} analysis of the 25 SDR proteins based on their associated sets of InterPro annotations, using the Jaccard similarity as the similarity measure. A distance threshold of $0.5$ was applied to the resulting dendrogram, which yielded four distinct clusters, one of which contained a single protein and was excluded from further consideration in our analysis. We then visualized the dendrogram generated by the \textit{hierarchical clustering} procedure alongside the protein sequence unit cluster visualization for each protein, which employs the PCA-based color assignment approach described above.

The assigned colors are largely conserved across all proteins, indicating that the representations of units within a given protein family are generally similar. Across these proteins, 11 distinct unit clusters are identified, with clusters 917, 670, and 281 being the most prevalent. Clusters 917 and 670 share the same three most frequent GO annotations: oxidoreductase activity, acting on the CH-OH group of donors (GO:0016614); oxidoreductase activity, acting on the CH-OH group of donors, NAD or NADP as acceptor (GO:0016616); and fatty acid synthase activity (GO:0004312). By contrast, the top three GO annotations for cluster 281 are fatty acid synthase activity (GO:0004312); oxidoreductase activity, acting on the CH-CH group of donors, NAD or NADP as acceptor (GO:0016628); and oxidoreductase activity, acting on the CH-CH group of donors (GO:0016627). Another cluster of interest is cluster 71, with GO terms hydro-lyase activity (GO:0016836), oxidoreductase activity, acting on the CH-OH group of donors, NAD or NADP as acceptor (GO:0016616), and O-acyltransferase activity (GO:0008374). The remaining clusters occur less frequently and consequently occupy a smaller fraction of the protein sequences.

There are 17 proteins assigned to the blue group. With the exception of \textit{1AHH}, which appears to represent a misassigned unit cluster, all of these proteins share similar unit cluster compositions. Looking at the GO annotations of these proteins, we see that this cluster predominantly contains SDRs that catalyze reactions involving a $C=O$ bond, consistent with the GO terms of the dominant clusters 917 and 670 utilized in this group. Although these two clusters share similar GO terms, their separation, with cluster 917 localized to the N-terminal region and cluster 670 to the C-terminal region, suggests a structural and functional distinction within the protein. This distinction aligns with the literature, which describes the N-terminal region as the cofactor-binding domain and the C-terminal region as the substrate-binding domain \cite{kallberg_sdr}. It is also noteworthy that, based on their GO annotations, \textit{1DIR} acts on $CH\text{-}NH$ groups and \textit{2FWM} acts on $CH\text{-}CH$ groups. This observation suggests that the use of InterPro-based annotations does not always yield functionally homogeneous or strictly distinct protein clusters, as these two proteins were grouped with enzymes acting on $CH\text{-}OH$.

In the red group, the proteins with PDB identifiers \textit{4JQC, 5CFZ, 1W73, 4FC7}, and \textit{1YXM} are observed. According to the GO annotations of these proteins, these correspond to SDRs that catalyze reactions involving a $C=C$ double bond. All but one of these proteins contain Cluster 281, which is associated with oxidoreductase activity acting on CH–CH groups. Consequently, for this group, the functional assignments generated by our method are consistent with the corresponding InterPro annotations.

The final green group corresponds to the human protein sepiapterin reductase (SPR). The two PDB chains shown represent the same protein bound to different inhibitors. Both structures contain Cluster 71; however, in this particular case, the associated GO terms do not accurately reflect the protein’s function. Furthermore, the differences in their unit cluster compositions, for instance, the presence of an additional cluster in \textit{4HWK} relative to \textit{4Z3K}, further illustrate the sensitivity of PUFFIN to structural variations induced by ligand binding.

Overall, our analysis of the SDR family demonstrates that PUFFIN can detect functionally relevant signals, such as detailed characteristics of oxidoreductase activity, and assign functional units accordingly. The presence of some incorrect unit cluster compositions and GO term assignments indicates that there is still room for improvement; however, this may be inherently challenging due to the noisy and potentially inconsistent functional annotations in InterPro.

\end{document}